\documentclass[aps,pre,showpacs,showkeys,twocolumn,floatfix,superscriptaddress]{revtex4}

\usepackage{amsmath} \usepackage{amsfonts} \usepackage{amssymb}
\usepackage{graphicx}
\usepackage{dcolumn}
\usepackage{bm}
\usepackage{color}

\begin{document}


\title{Discrete breathers in a nonlinear electric line: \\
       Modeling, Computation and Experiment}

\author{F. Palmero}
\email[Corresponding author. Electronic address:]{palmero@us.es}
\affiliation{Nonlinear Physics
Group. Escuela T\'{e}cnica Superior de Ingenier\'{\i}a
Inform\'{a}tica. Departamento de F\'{\i}sica Aplicada I. Universidad de
Sevilla. Avda.  Reina Mercedes, s/n. 41012-Sevilla (Spain)}

\author{L. Q. English}
\affiliation{Department of Physics and Astronomy, Dickinson College, Carlisle, Pennsylvania 17013, USA}

\author{J.\ Cuevas}
\affiliation{Grupo de F\'{\i}sica No Lineal. Departamento de
F\'{\i}sica Aplicada I, Escuela Polit\'ecnica Superior. Universidad de Sevilla,
C/ Virgen de \'{A}frica, 7. 41011-Sevilla, Spain}

\author{R.\ Carretero-Gonz\'alez}
\affiliation{Nonlinear Dynamical Systems Group,%
\footnote{\texttt{URL}: http://nlds.sdsu.edu}%
Department of Mathematics and Statistics, and Computational Science
Research Center, San Diego State University, San Diego CA, 92182-7720, USA}

\author{P.G.\ Kevrekidis}
\affiliation{Department of Mathematics and Statistics, University
of Massachusetts, Amherst, Massachusetts 01003-4515, USA}

\date{\today}

\begin{abstract}

We study experimentally and numerically the existence and
stability properties of discrete breathers in a periodic
nonlinear electric line.
The electric line is composed of single cell nodes, containing
a varactor diode and an inductor, coupled together in a periodic
ring configuration through inductors and driven uniformly by a
harmonic external voltage source.
A simple model for each cell is proposed by using a
nonlinear form for the varactor characteristics through
the current and capacitance dependence on the
voltage.
For an electrical line composed of 32 elements, we
find the regions, in driver voltage and frequency,
where $n$-peaked breather solutions exist and
characterize their stability. The results are
compared to experimental measurements with good
quantitative agreement.
We also examine the spontaneous formation
of $n$-peaked breathers through modulational instability of
the homogeneous steady state. The competition between
different discrete breathers seeded by the modulational
instability eventually leads to stationary $n$-peaked
solutions whose precise locations is seen to sensitively
depend on the initial conditions.

\end{abstract}

\pacs{05.45.Yv, 63.20.Pw, 63.20.Ry}
\keywords{Localized modes; Solitons; Discrete
Breathers}

\maketitle
%


\section{Introduction}
Nonlinear physics of discrete systems has witnessed enormous development in the past years. In particular, a great deal of interest has been paid to the existence and properties of intrinsic localized modes (ILMs), or discrete breathers, which result from the combination of nonlinearity and spatial discreteness.
These spatially localized states have been observed in a wide variety of different systems \cite{camp04,Flach2008}. They were originally suggested as
excitations of   anharmonic nonlinear lattices \cite{sietak}, but
the rigorous proof of their persistence under general
conditions~\cite{macaub} led to their investigation
in a diverse host of applications. These include, among others,
antiferromagnets~\cite{Schwarz1999}, charge-transfer solids~\cite{Swanson1999},
photonic crystals~\cite{photon}, superconducting Josephson junctions~\cite{orlando1}, micromechanical cantilever arrays~\cite{sievers}, granular
crystals~\cite{theo} and
 biopolymers~\cite{Xiepeyrard}.
More recently, the direct manipulation and control of such
states has been enabled through suitable experimental
techniques \cite{control}.

Despite the tremendous strides made in this field,
the relevant literature, nevertheless, often appears to be quite sharply
divided between theory and experiment. Frequently, experimental studies do not
capture the dynamics in enough detail to facilitate an exact comparison with
theoretical studies. At other times, the theoretical models are not refined
enough (or lack the inclusion of nontrivial experimental factors some of which may be difficult to quantify precisely) to make
quantitative contact with the experimental results. Our system
---a macroscopic electrical lattice in which solitons have a
time-honored history~\cite{marq94,remoi}--- is, arguably,
ideally suited for this kind of cross-comparison: the lattice dynamics
can be measured fully in space and time, and the physical properties of
individual unit cells of the lattice can be characterized in enough
detail to allow for the construction of effective models.

In this paper, we present a detailed study of discrete breathers in an
electric lattice in which ILMs have been experimentally observed
\cite{Lars07, Lars08, Lars10}. We propose a  theoretical
model which allows us to systematically
study their existence, stability and properties,
and to compare our numerical findings with experimental results.
We demonstrate good agreement not only at a qualitative but also at
a quantitative level between theory and experiment.
The presentation of our results
is organized as follows. In the next section we study
the characteristics (intensity and capacitance curves versus voltage)
of the varactor, the nonlinear circuit element, in order to develop the relevant model for the electrical
unit cell. The results for the single cell are validated through the
comparison of its resonance curves for different driving strengths.
We also derive the equations describing the entire electrical line.
In Sec.~\ref{Sec:numexp} we study the existence and stability properties
of $n$-peaked breathers for $n=1,2,3$ in the driving frequency and voltage
parameter space. The numerical results are compared to the experimental
data with good quantitative agreement.
We also briefly study the spontaneous formation of $n$-peaked
breathers from the modulational instability of the homogeneous steady
state. We observe that the location and number of the final
peaks depends sensitively on the initial conditions.
Finally, in Sec.~\ref{Sec:conclu}, we conclude our manuscript
and offer some suggestions for possible avenues of
further research.

\section{Theoretical Setup}

\begin{figure}
\begin{center}
\includegraphics[scale=0.45]{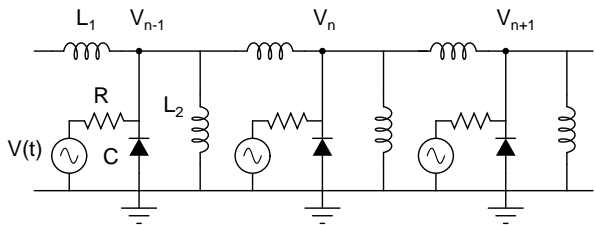}
\end{center}
\caption{Schematic circuit diagram of the electrical transmission line.}
\label{line}
\end{figure}

Our system consists on an electric line as represented in Fig.~\ref{line}.
This line can be considered as a set of single cells, each one composed of
a varactor diode (NTE 618) and an inductor  $L_2=330$~$\mu$H, coupled
through inductors $L_1=680$~$\mu$H. Each unit cell or node is driven via a
resistor, $R=10$~k$\Omega$, by a sinusoidal voltage source $V(t)$ with
amplitude $V_d$ and frequency $f$. In experiments a set of 32 elements have
been used, with a periodic ring structure (the last element is connected to
the first one), and measurements of voltages $V_n$ have been recorded.
Related to the voltage source, we have considered amplitudes from $V_d=1$~V
to $V_d=5$~V and frequencies from $f=200$~kHz to $f=600$~kHz.

\begin{figure}[ht]
\begin{center}
\includegraphics[scale=0.5]{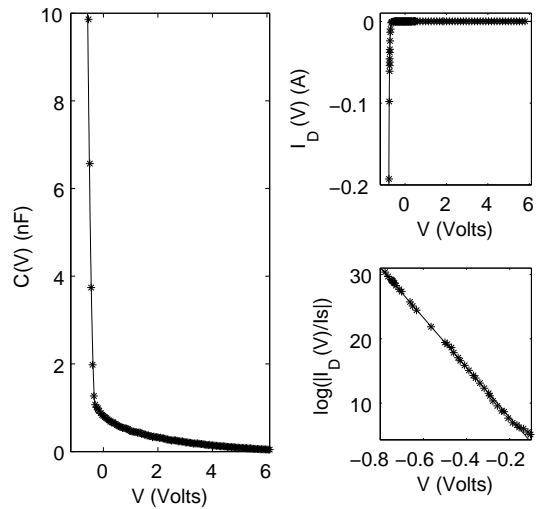}
\end{center}
\caption{Experimental data ($\ast$) and numerical approximation thereof
(continuous line) corresponding to $C(V)$ and $I_D(V)$ (linear and
semi--log plots) for the nonlinear varactor.}
\label{expe}
\end{figure}

In order to propose a set of equations to characterize the electrical line,
we have used circuit theory and Kirchhoff's rules; the main  challenge has
been to describe appropriately each element. In general, resistance and
inductors are inherently imperfect impedance components, i.e., they have
series and parallel, reactive, capacitive and resistive elements. Moreover,
due to the commercial nature of the elements, manufactured components are
subject to tolerance intervals, and the resultant small spatial
inhomogeneity introduces some additional uncertainty. We have quantified the
spatial inhomogeneity by separately
measuring all lattice components. The diode
capacitance was found to vary by 0.3 percent (standard deviation), whereas
the inductors both exhibited a 0.5 percent variation. Additional factors that may contribute slightly to inhomogeneities are wire inductances as well as load and contact resistances. The varactors (diodes)
we use are typically intended for AM receiver electronics and tuning
applications. As described later, we characterize this lattice element in
more detail, since it is the source of the nonlinearity in the lattice.

As a guiding principle, we are aiming to construct a model which is as simple
as possible, with a limited number of parameters whose values are
experimentally supported, but one which is still able to reproduce the main
phenomenon, namely nonlinear localization and the formation of discrete
breathers. With this balance in mind, we proceed as follows.

In the range of frequencies studied it is a good approximation to describe the
load resistance as a simple resistor, neglecting any capacitive or inductive
contribution. Also, we have performed experimental measures of the varactor
characteristics. This experimental data shows that it can be modeled as a
nonlinear resistance in parallel with a nonlinear capacitance, where the
nonlinear current $I_D(V)$  is given by
\begin{equation}
 I_D(V)= -I_s \exp(-\beta V),
\end{equation}
%
%
where $\beta=38.8$~V$^{-1}$ and $I_s=1.25 \times 10^{-14}$ A (we consider
negative voltage  when the varactor is in direct polarization), and its
capacitance as
\begin{equation}
 C(V)=
\begin{cases} \begin{matrix}
C_{v}+C_{w}(V')+  C(V')^2  & \mbox{if} \quad V \leq V_{c}, \\[2.0ex]
  C_0 e^{-\alpha V} & \mbox{if} \quad V > V_{c},
\end{matrix} \end{cases}
\end{equation}
where $V'=(V-V_c)$, $C_0=788$~pF, $\alpha=0.456$~V$^{-1}$,
$C_{v}=C_0\exp(-\alpha V_{c})$,  $C_{w}=-\alpha C_{v}$ (the capacitance and
its first derivative are continuous in $V=V_{c}$), $C=100$~nF and
$V_{c}=-0.28$~V. In Fig.~\ref{expe} we present the experimental data and
their corresponding numerical approximations for $I_D(V)$ and $C(V)$, where a
good agreement between the two can be observed.

\begin{figure}[ht]
\begin{center}
\includegraphics[scale=0.8]{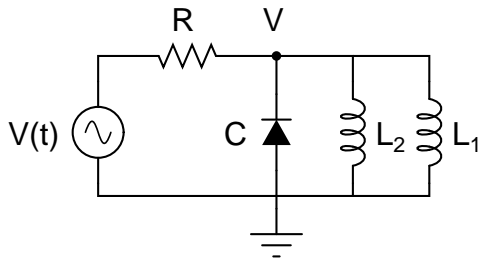}
\end{center}
\caption{Single cell element model.}
\label{single}
\end{figure}

With respect to the inductors, in the range of frequencies considered,
capacitive effects are negligible, but they possess a small dc ohmic
resistance which is around $2 \Omega$.  The inductors and the varactor are a source of damping in the
ac regime, and these contributions  must be taken into account. However, we
have no  manufacturer data related to dissipation parameters, and it is
difficult to measure them experimentally.
In order to introduce these effects,  we will model dissipation
phenomenologically by means of a global term given by a resistance $R_l$,
which appears in each unit cell in parallel with $L_2$; to determine its
value, we have studied experimentally a single element as shown in
Fig.~\ref{single}. In this way, we will consider the inductors themselves as ideal
elements.

Using basic circuit theory, the single element is described by the equations:
\begin{eqnarray}
 \frac{d v}{d \tau} &=& \frac{1}{c(v)}\left[\frac{\cos(\Omega \tau)}{R C_0 \omega_0} \right.- \\
 & & \left. \frac{1}{\omega_0C_0} \left(\frac{1}{R_l}+\frac{1}{R}\right) v+(y-i_D)\right], \nonumber \\[2.0ex]
 \frac{d y}{d \tau}& = & -\left(1+\frac{L_2}{L_1}\right) v,
\end{eqnarray}
where dimensionless variables have been used:  $\tau= \omega_0 t$,  $i_D=I_D/(\omega_0C_0V_d)$, $v=V/V_d$,  the dimensionless voltage at point $A$, $c(v)=C(V)/C_0$, $\Omega=\omega/\omega_0$, and $\omega_0=1/\sqrt{L_2C_0}$; $y$ represents the normalized current through the inductors.

\begin{figure}
\begin{center}
\includegraphics[scale=0.5]{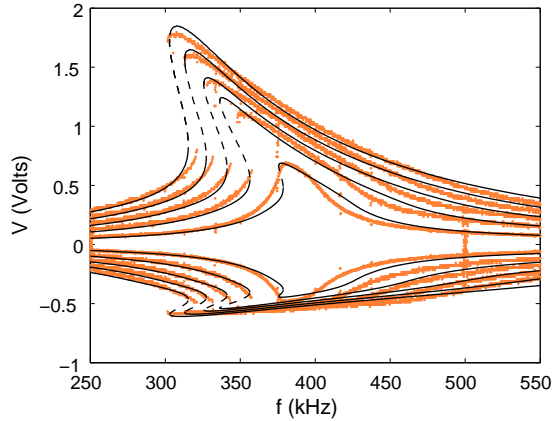}
\end{center}
\caption{(Color online).
Nonlinear resonance curves corresponding to a single element.
Dots (orange [grey]) correspond to experimental data while the continuous and
dashed black lines correspond, respectively, to stable and unstable
numerical solutions. From bottom to top: $V_d =1, 2 ,3, 4$ and 5 V.}
\label{ress}
\end{figure}

We can generate theoretical nonlinear resonance curves and, comparing with
experimental data, select the optimal dissipation
parameter value $R_l$. Results are summarized in Table \ref{rl}, and the
comparison between theoretical and experimental data is shown in
Fig.~\ref{ress}. Also, we consider a small frequency shift of $18$~kHz in
numerical simulations to match  the resonance curves. This effect may
originate from some small capacitive and/or inductive contributions that
we have not previously taken into account.

\begin{table}[ht]
\begin{center}
\begin{tabular}{|c|c|c|c|c|c|c|c|}
\hline
$V_d$ (Volts) &  1 & 2 & 3 & 4 & 5 \\
\hline
 $R_l$ ($\Omega$) & 15000 & 10000 & 6000 & 5000 & 4500 \\
 \hline
 \end{tabular}
  \end{center}
  \caption{Values of the resistance $R_l$ corresponding to different
voltage amplitudes for the driving source $V_d$.}
 \label{rl}
\end{table}

Thus, with a consistent set of parameter values, and using again elementary
circuit theory, we describe the full electric line of $N$ coupled single cell
elements by the following system of coupled ordinary differential equations:
\begin{eqnarray}
 c(v_n) \frac{d v_n}{d \tau} & = & y_n-i_D(v_n)+ \frac{\cos(\Omega \tau)}{RC_0\omega_0} -
\nonumber  \\ & &
\left(\frac{1}{R_l}+\frac{1}{R}\right) \frac{v_n}{\omega_0 C_0} \nonumber, \\[2.0ex]
 \frac{d y_n}{d\tau} & = & \frac{L_2}{L_1}\left(v_{n+1}+v_{n-1}-2v_n\right) -v_n,  \\ \nonumber
\label{eqline}
\end{eqnarray}
where all magnitudes are in dimensionless units.
Within this nonlinear dynamical lattice model, our waveforms of interest,
namely the discrete breathers, are calculated as fixed points of the map
\begin{eqnarray}
\left[\begin{array}{c} v_n(0) \\[1.0ex]
 \frac{d v_n}{d\tau}(0) \\[1.0ex]
  y_n(0) \\[1.0ex]
 \frac{d y_n}{d\tau}(0) \end{array} \right] \rightarrow
\left[\begin{array}{c} v_n(T) \\[1.0ex]
 \frac{d v_n}{d\tau}(T) \\[1.0ex]
  y_n(T) \\[1.0ex]
 \frac{d y_n}{d\tau}(T) \end{array} \right],
\end{eqnarray}
where $T=1/f$ is the temporal period of the breather.
In order to study the linear stability of discrete breathers, we introduce a small perturbation $(\xi_n,\eta_n)$ to a given solution $(v_{n0},y_{n0})$ of Eq.~(\ref{eqline}) according to $v_n=v_{n0}+\xi_n$, $y_n=y_{n0}+\eta_n$.
Then, the equations satisfied to first order by $(\xi_n,\eta_n)$ are:
\begin{eqnarray}\label{eqstab}
    c(v_{n0}) \frac{d \xi_n}{d \tau} &=& \eta_n-\Gamma(v_{n0};t)\xi_n \nonumber \\[2.0ex]
    \frac{d \eta_n}{d \tau} &=& \frac{L_2}{L_1}\left(\xi_{n+1}+\xi_{n-1}-2\xi_n\right)-\xi_n
\end{eqnarray}
with $\Gamma(v_{n0};t)$ being
\begin{eqnarray}
    \Gamma(v_{n0};t) & = & \frac{d i_D(v_{n0})}{d v_{n0}}+\left(\frac{1}{R_l}+\frac{1}{R}\right)\frac{1}{\omega_0 C_0}+ \nonumber \\[1.0ex]
 \nonumber & & \frac{d \ln[c(v_{n0})]}{d v_{n0}} \left[y_{n0}-i_D(v_n0)+\frac{\cos(\Omega\tau)}{RC_0\omega_0}\right. \\[1.0ex]
 \nonumber & & \left.- \left(\frac{1}{R_l}+\frac{1}{R}\right)\frac{v_{n0}}{\omega_0 C_0}\right].
\end{eqnarray}
To identify the orbital stability of the relevant solutions, a Floquet
analysis can be performed. Then, the stability
properties are given by the spectrum of the Floquet operator $\mathcal{M}$ (whose matrix representation is the monodromy) defined as:
\begin{equation}
    \left[\begin{array}{c} \xi_n(T) \\[1.0ex]
 \frac{d \xi_n}{d\tau}(T) \\[1.0ex]
 \eta_n(T) \\[1.0ex]
 \frac{d \eta_n}{d\tau}(T) \end{array} \right]=\mathcal{M}
    \left[\begin{array}{c} \xi_n(0) \\[1.0ex]
 \frac{d \xi_n}{d\tau}(0) \\[1.0ex]
 \eta_n(0) \\[1.0ex]
 \frac{d \eta_n}{d\tau}(0) \end{array} \right].
\end{equation}
The $4N\times4N$ monodromy eigenvalues $\lambda$ are called the {\em Floquet multipliers}. If the breather is stable, all the eigenvalues lie inside the unit circle.

\section{Numerical Computations \& Experimental Results}
\label{Sec:numexp}

Using Eq.~(\ref{eqline}), we have generated $n$-peak ILMs and determined
numerically their stability islands in (V$_d$, $f$) parameter space. We have
found that there exist overlapping regions where two, or several, of these
$n$-peak configurations exist and are stable. Therefore, the long term
dynamics in these
regions is chiefly dependent on initial conditions. However, determining
precisely the basins of attraction of each configuration is not possible
because of the high dimensionality of the problem.

\begin{figure}
\begin{center}
\begin{tabular}{ccc}
\includegraphics[scale=0.4]{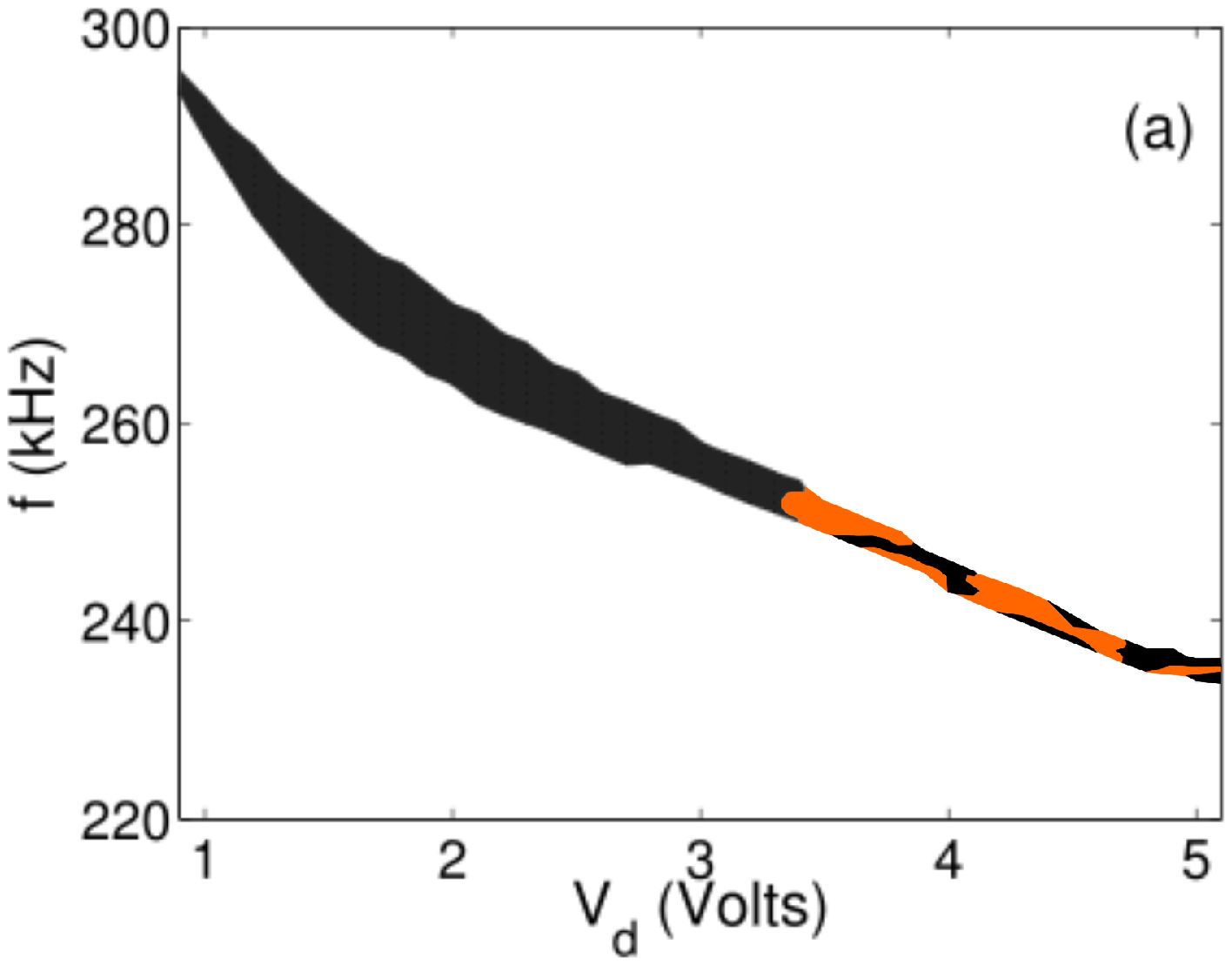} \\
\includegraphics[scale=0.4]{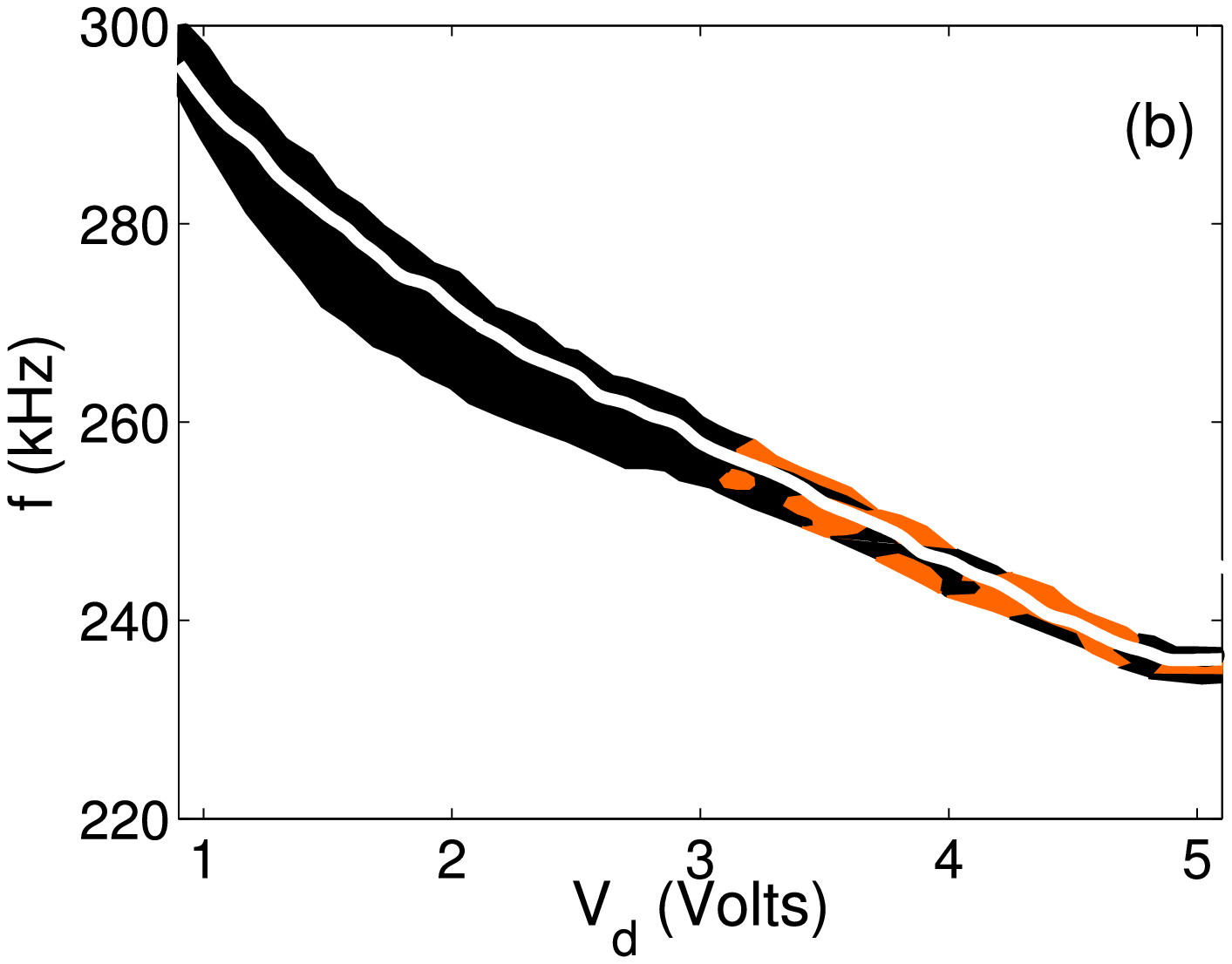} \\
\includegraphics[scale=0.4]{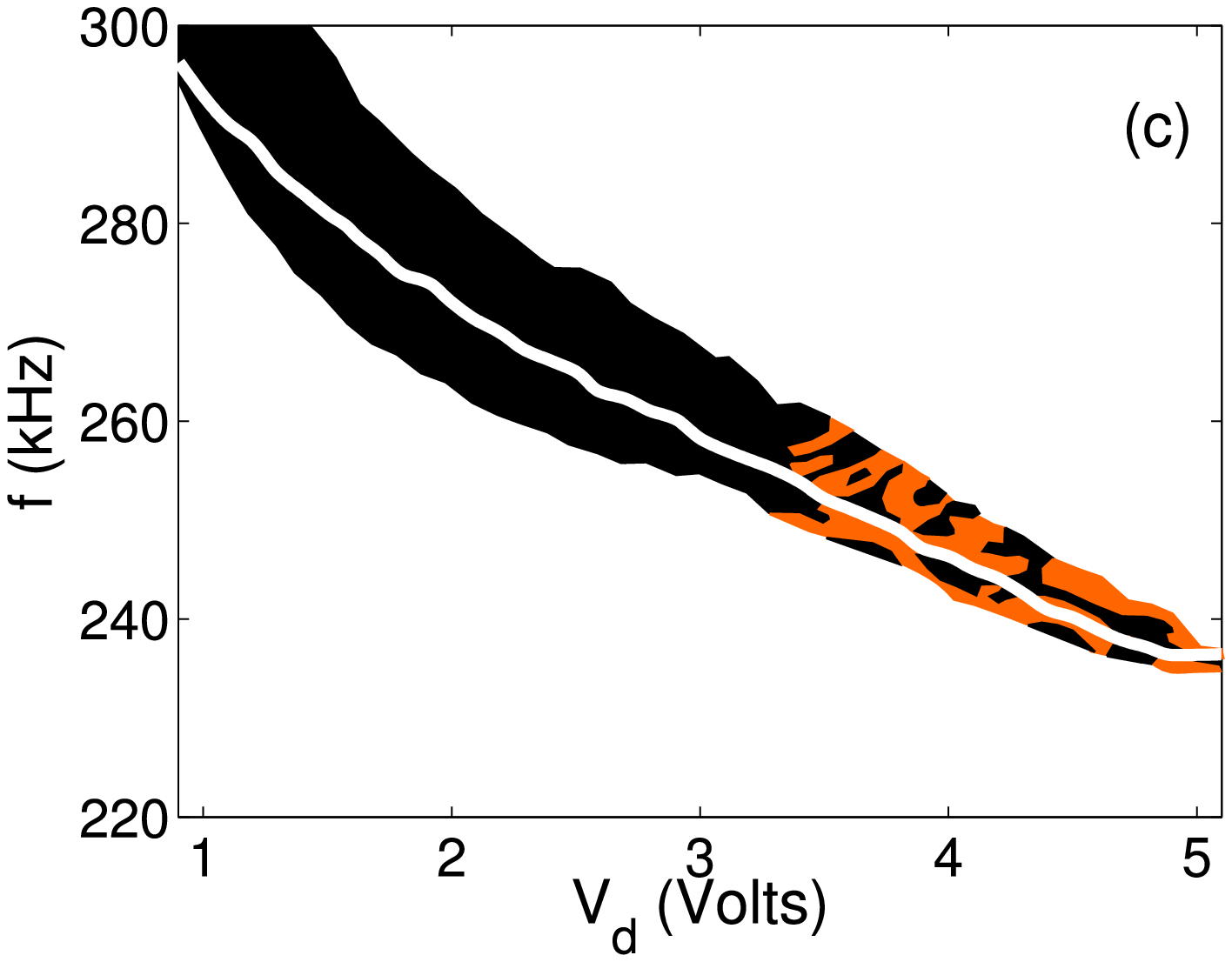}
\end{tabular}
\end{center}
\caption{(Color online).
Existence islands for $n$-peak breathers. The white line denotes the
threshold where the homogeneous state becomes unstable (above
the line). Black areas correspond to stable solutions and orange (grey) areas to unstable solutions. (a) one-peak breather (the homogeneous state threshold is located at the top border of the areas), (b) a family of two-peak breather and (c) a family of three-peak breathers.}
\label{RC1}
\end{figure}

Figure~\ref{RC1} shows the existence and stability diagrams for 1-, 2-
and 3-peak breathers. The values of $R_l$ are obtained through cubic
interpolation from Table \ref{rl}. Notice that  the lattice allows
different configurations of solutions of $n$-peak breathers for $n>1$
corresponding to the peaks centered at different locations, but
determining precisely the diagrams corresponding to all possible
configurations is not possible because of the size of the lattice.
Therefore, we have studied only breathers with peaks as far apart
as possible.

\begin{figure}
\begin{center}
\begin{tabular}{cc}
\includegraphics[scale=0.222]{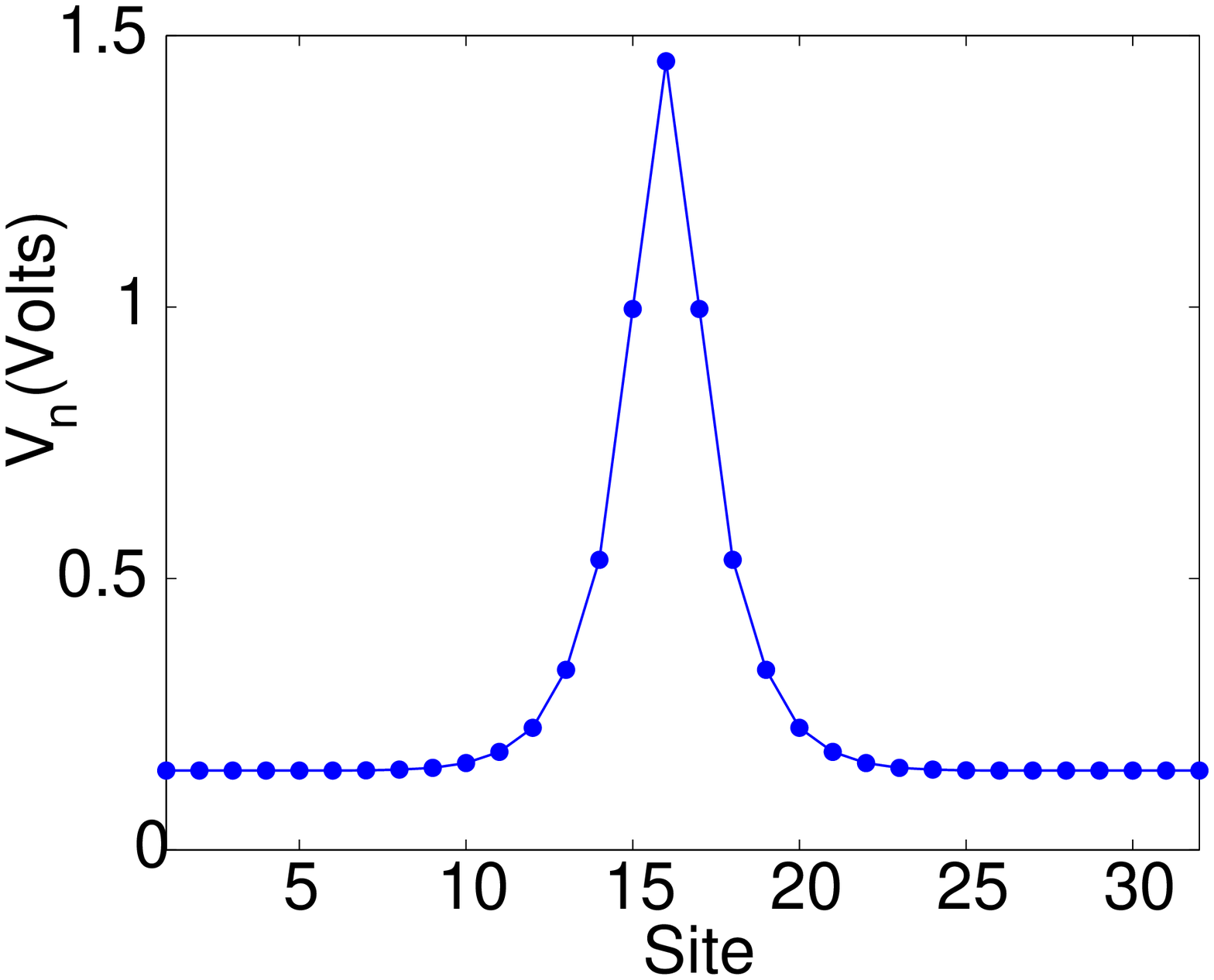} &
\includegraphics[scale=0.222]{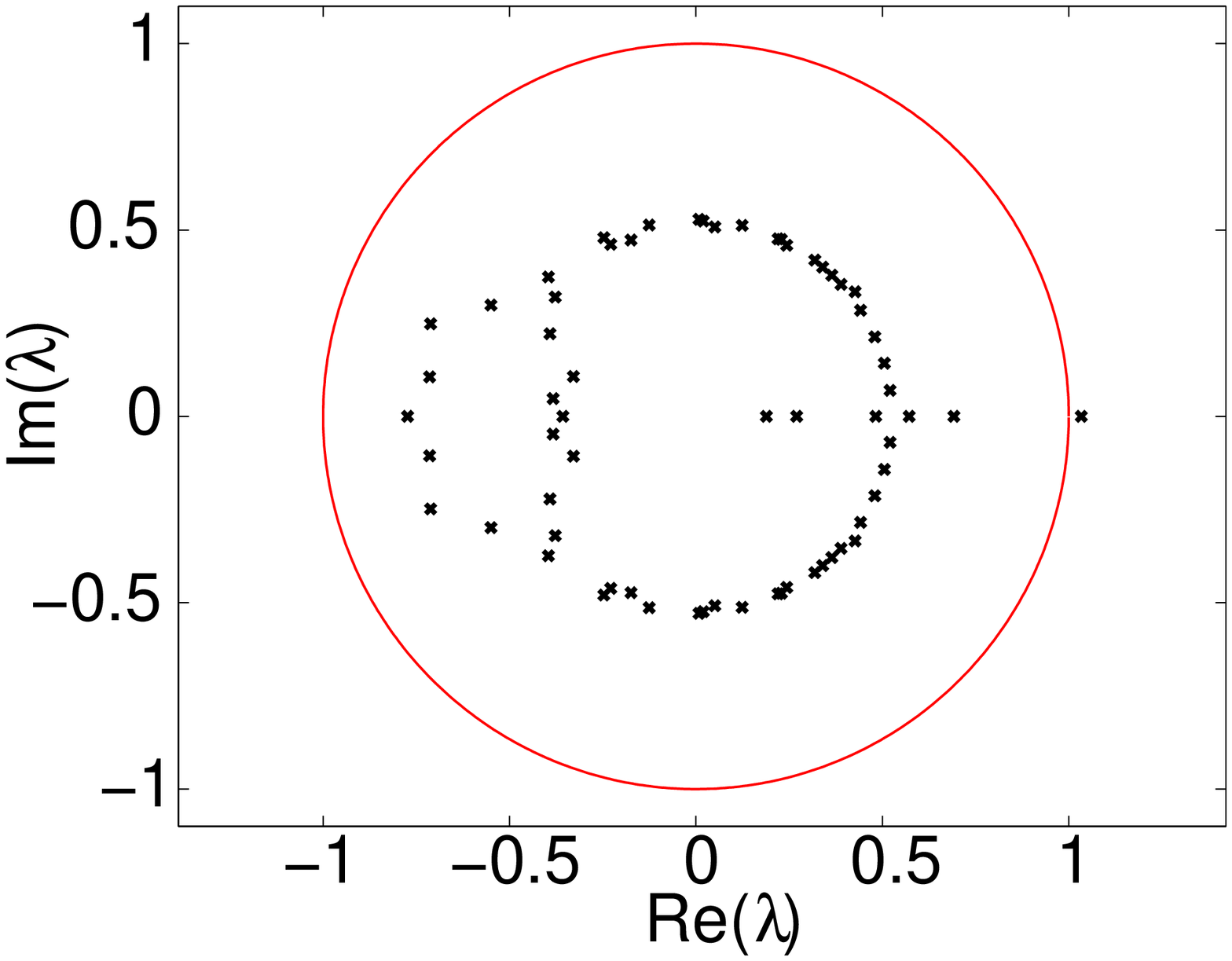} \\
\includegraphics[scale=0.222]{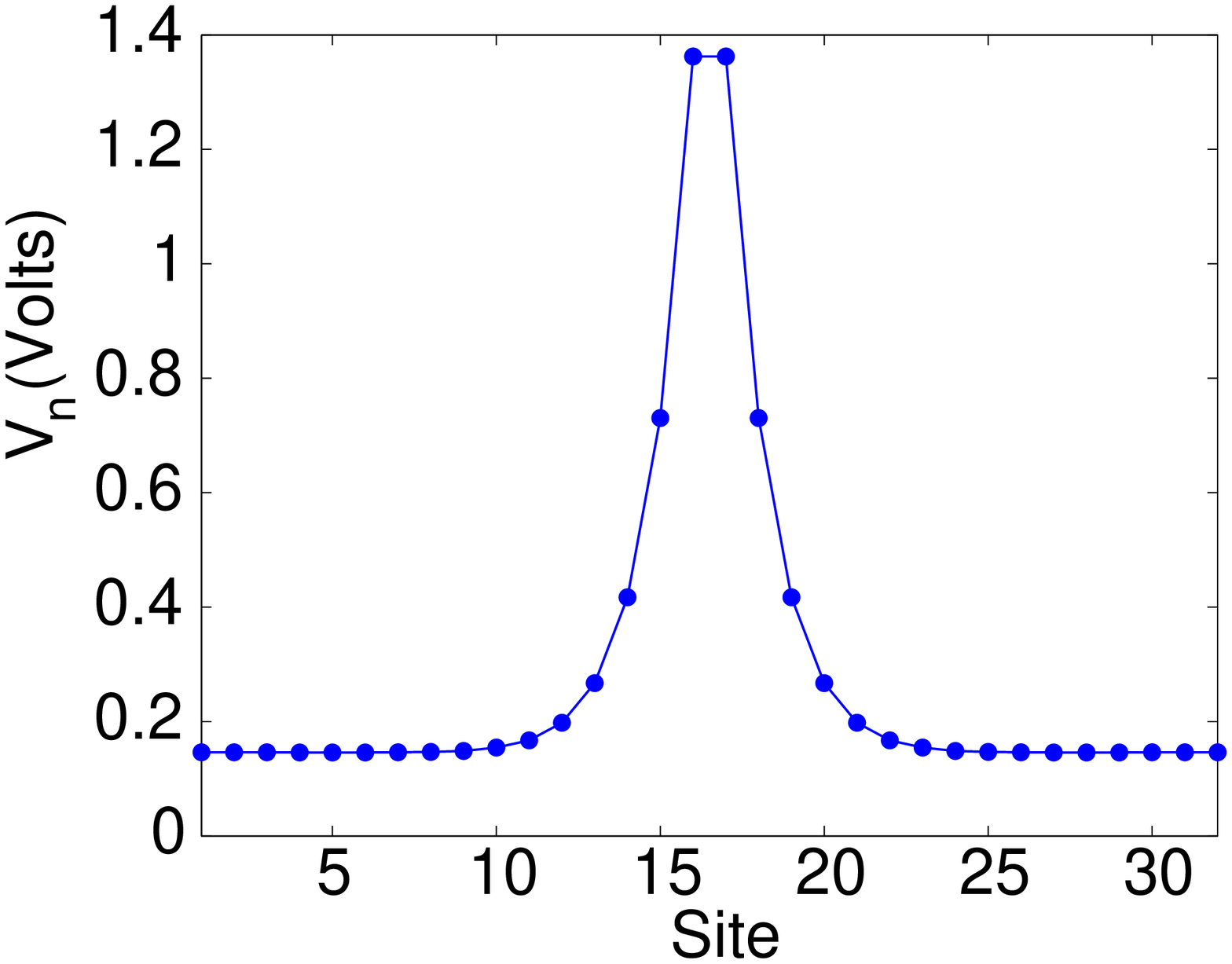} &
\includegraphics[scale=0.222]{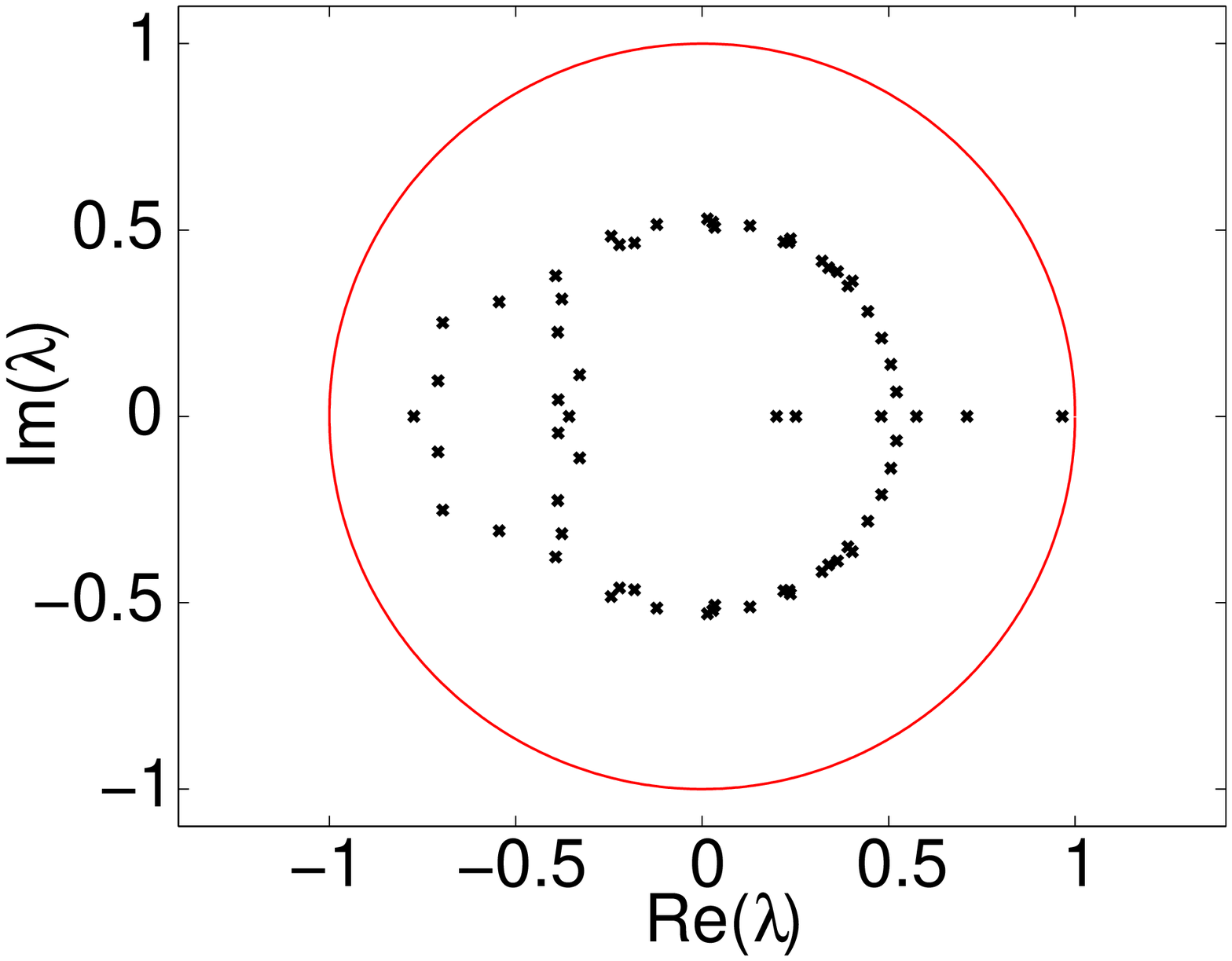} \\
\end{tabular}
\end{center}
\caption{(Color online).
Breather profiles (left column) and corresponding Floquet multiplier spectra
(right column) for $\tau=0$, $V_d=4$~V and $f=243.6$~kHz.
The top (bottom) panels correspond to the unstable (stable)
one-site (inter-site) breather.}
\label{examples}
\end{figure}

Remarkably, breathers are generally robust for $V_d\lesssim3$~V. Above that
critical value, the considered solutions may become unstable in
some ``instability windows'' (see orange [grey] regions). Those
instabilities, which are of exponential kind, typically lead an
onsite breather profile to deform into a stable
inter-site breather waveform. We have analyzed in more detail the dependence
of those instabilities for 1-peak breathers. The study for higher
peaked structure is cumbersome due to the increasing number of
inter-site structures that might arise. Figure~\ref{examples}
shows an example of 1-peaked on-site and inter-site breathers
together with their Floquet spectrum.

\begin{figure}
\begin{center}
\begin{tabular}{cc}
\multicolumn{2}{c}{on-site breather} \\
\includegraphics[scale=0.23]{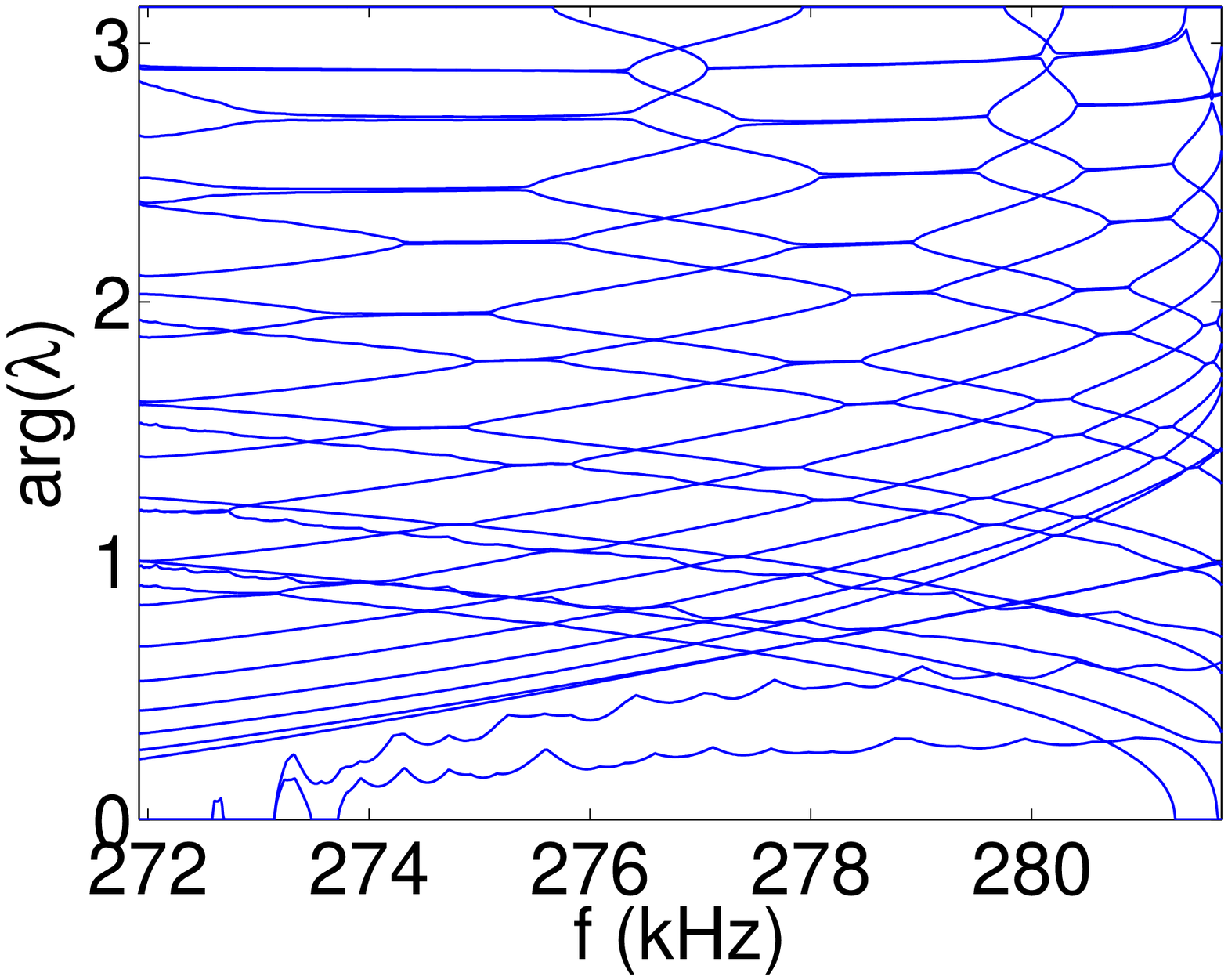} &
\includegraphics[scale=0.23]{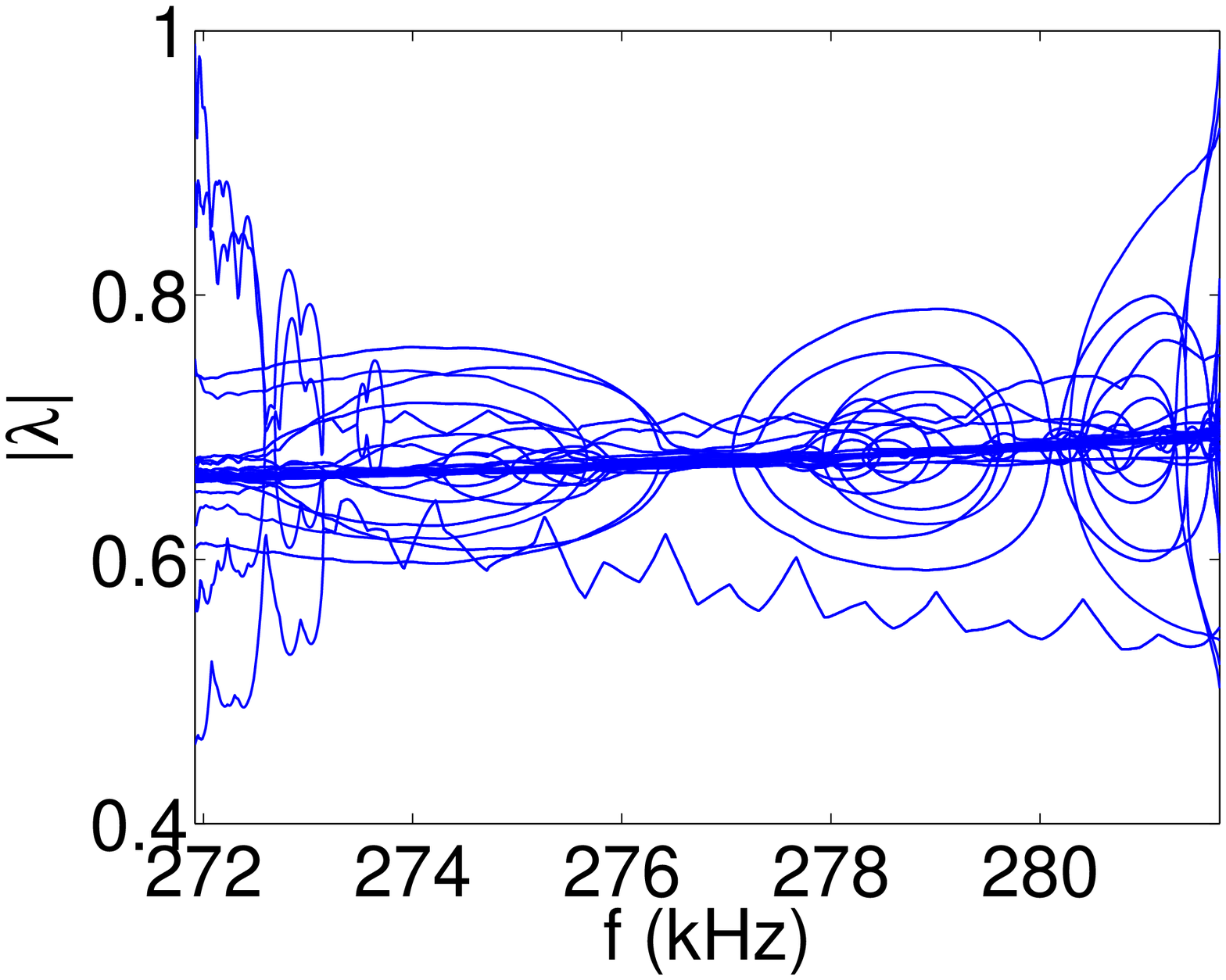} \\
\multicolumn{2}{c}{inter-site breather} \\
\includegraphics[scale=0.23]{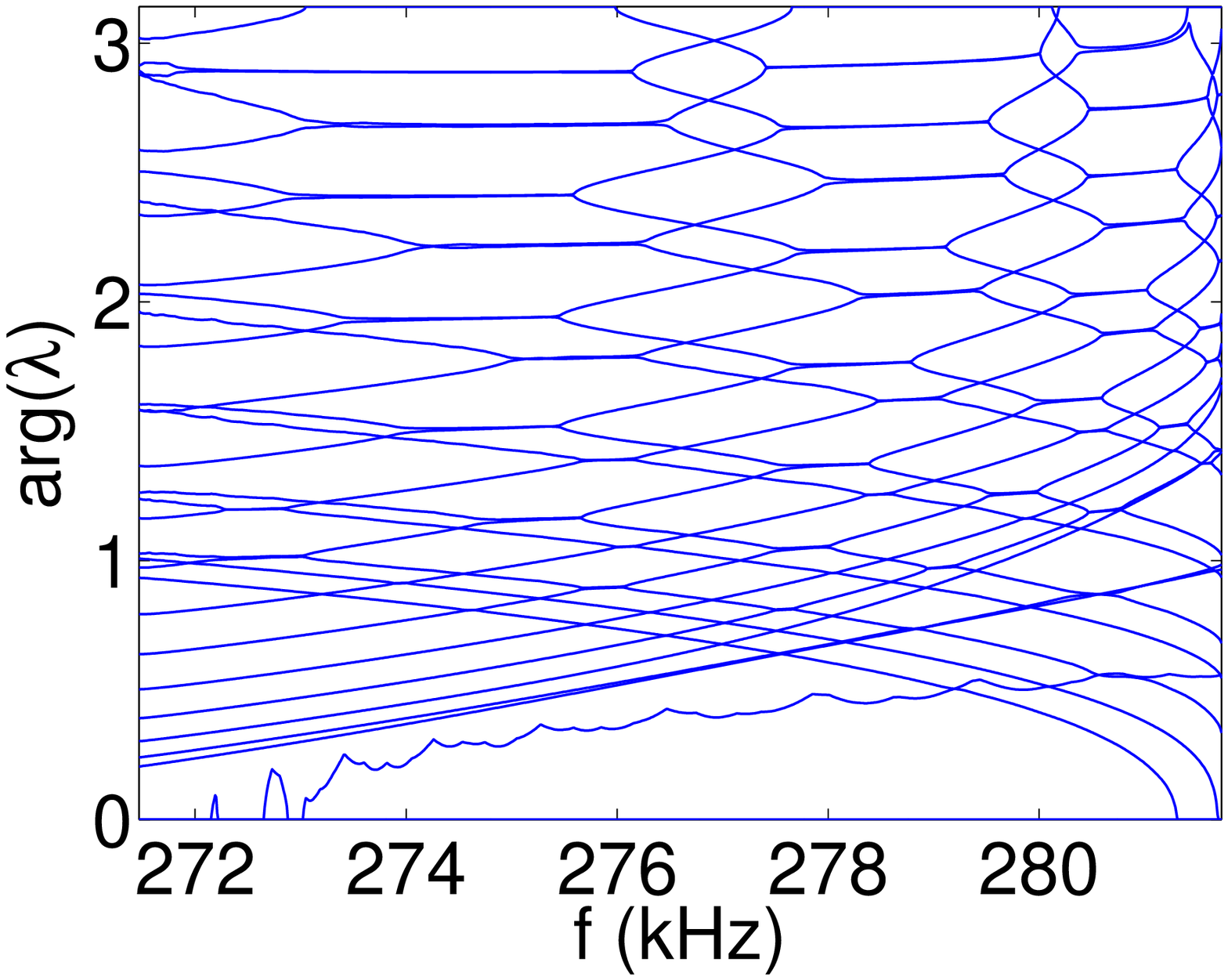} &
\includegraphics[scale=0.23]{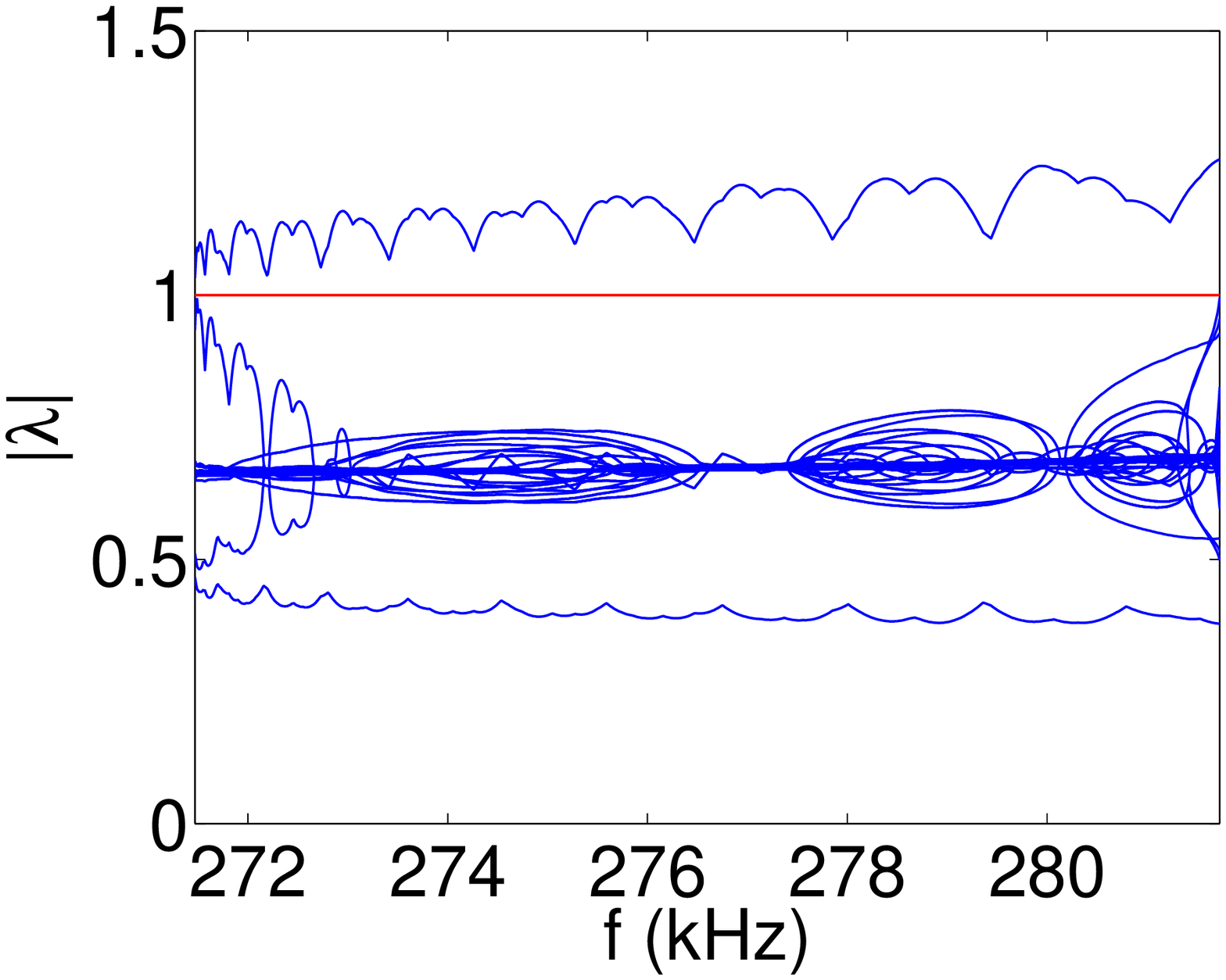} \\
\end{tabular}
\end{center}
\caption{(Color online).
Dependence of the Floquet multipliers on the frequency for $N=32$ cells and $V_d=1.5$~V.
The top (Bottom) row of panels corresponds
to the on-site (inter-site) breather.
The left and right column of panels depict, respectively, the argument and
magnitude of the Floquet multipliers. The horizontal (red)  line in the
spectra depicts the stability threshold.}
\label{exchange1}
\end{figure}

\begin{figure}
\begin{center}
\begin{tabular}{cc}
\multicolumn{2}{c}{on-site breather} \\
\includegraphics[scale=0.23]{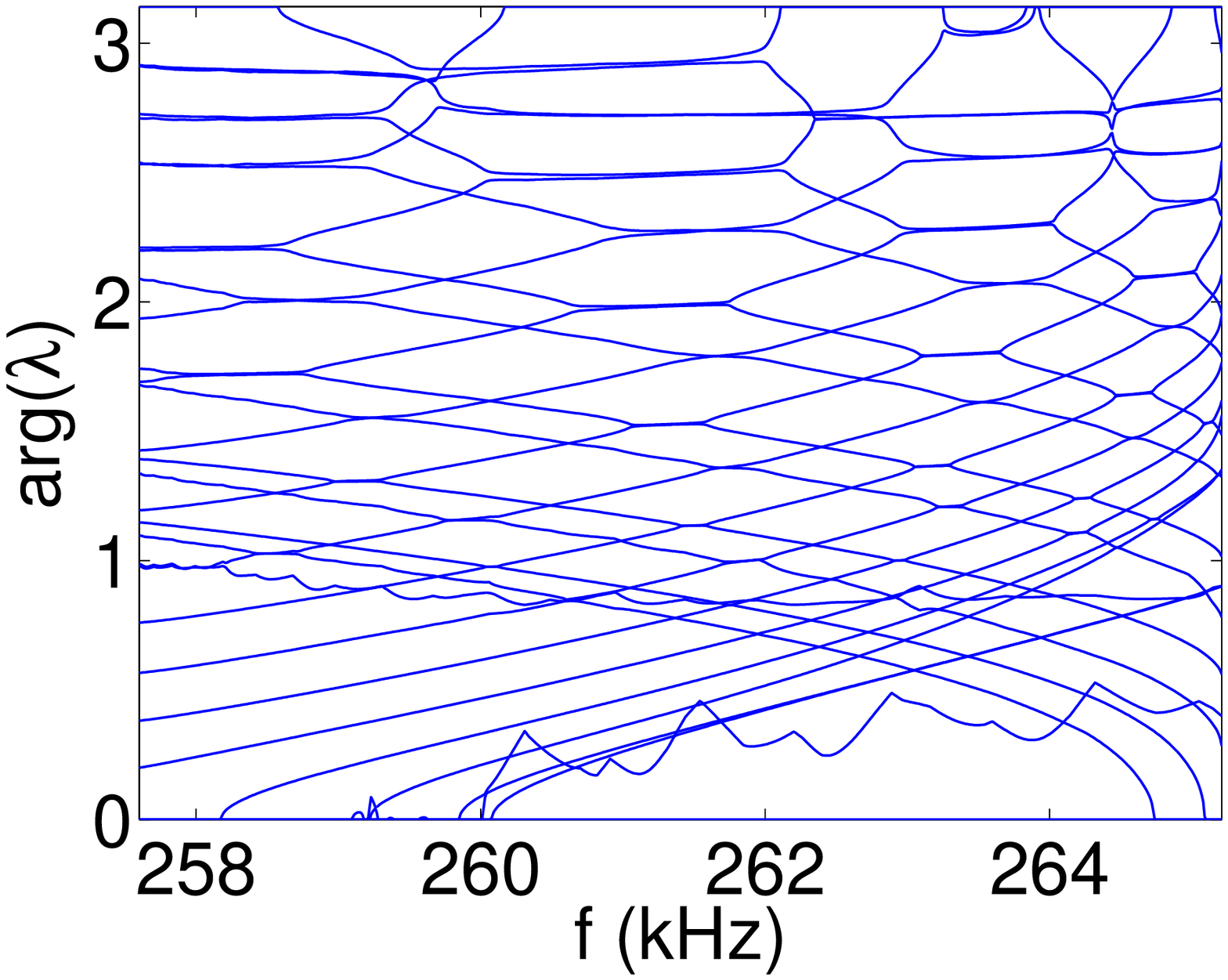} &
\includegraphics[scale=0.23]{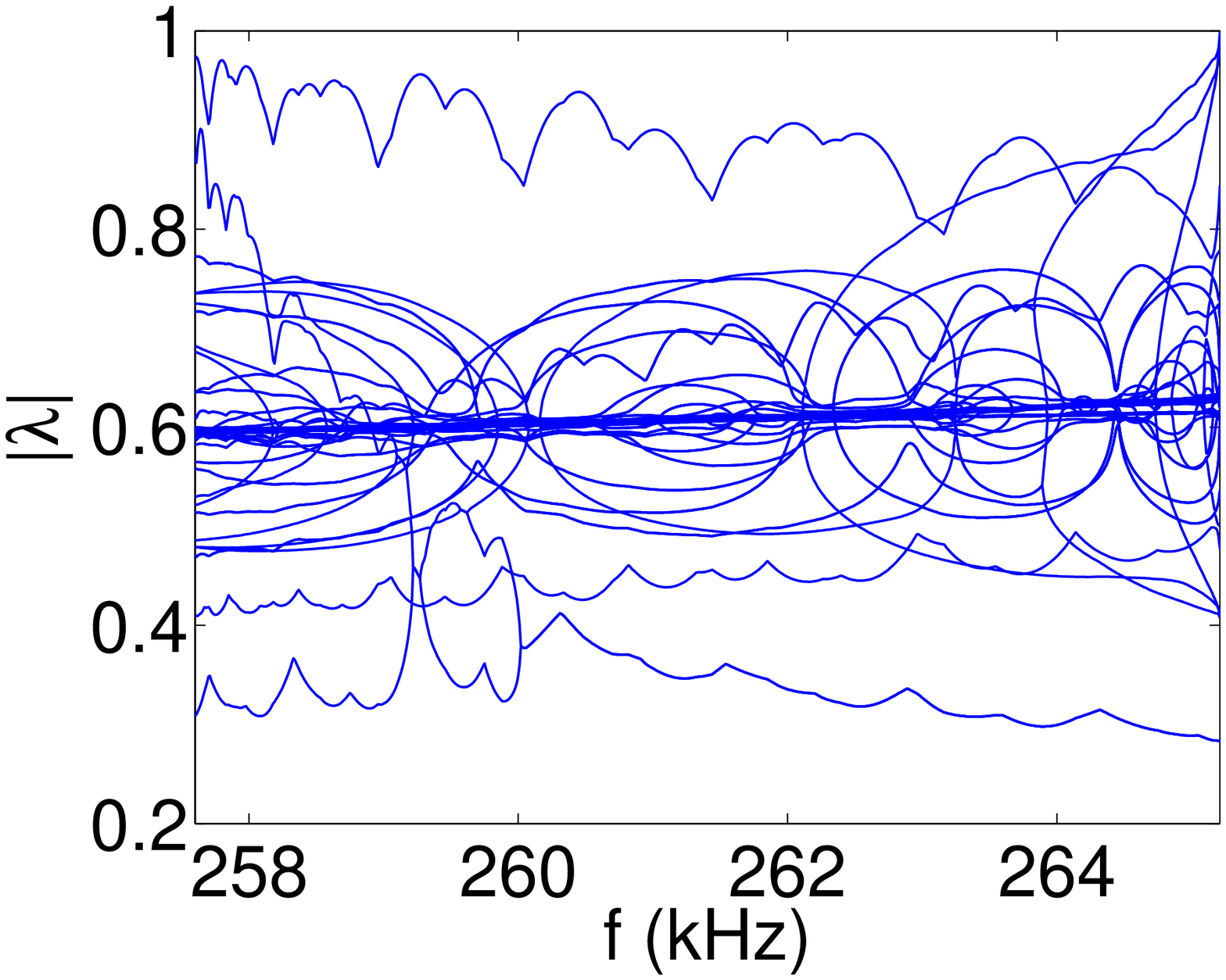} \\
\multicolumn{2}{c}{inter-site breather} \\
\includegraphics[scale=0.23]{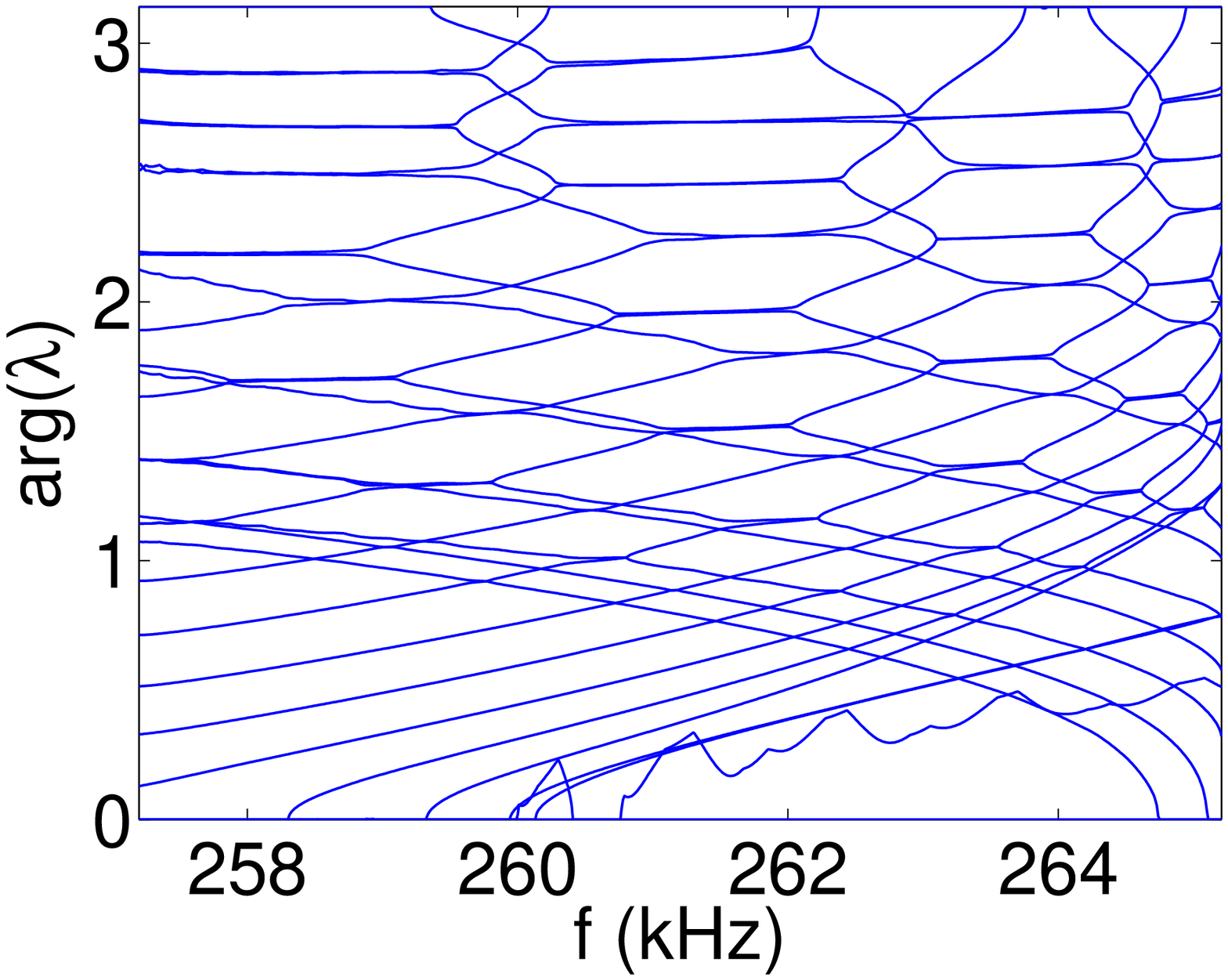} &
\includegraphics[scale=0.23]{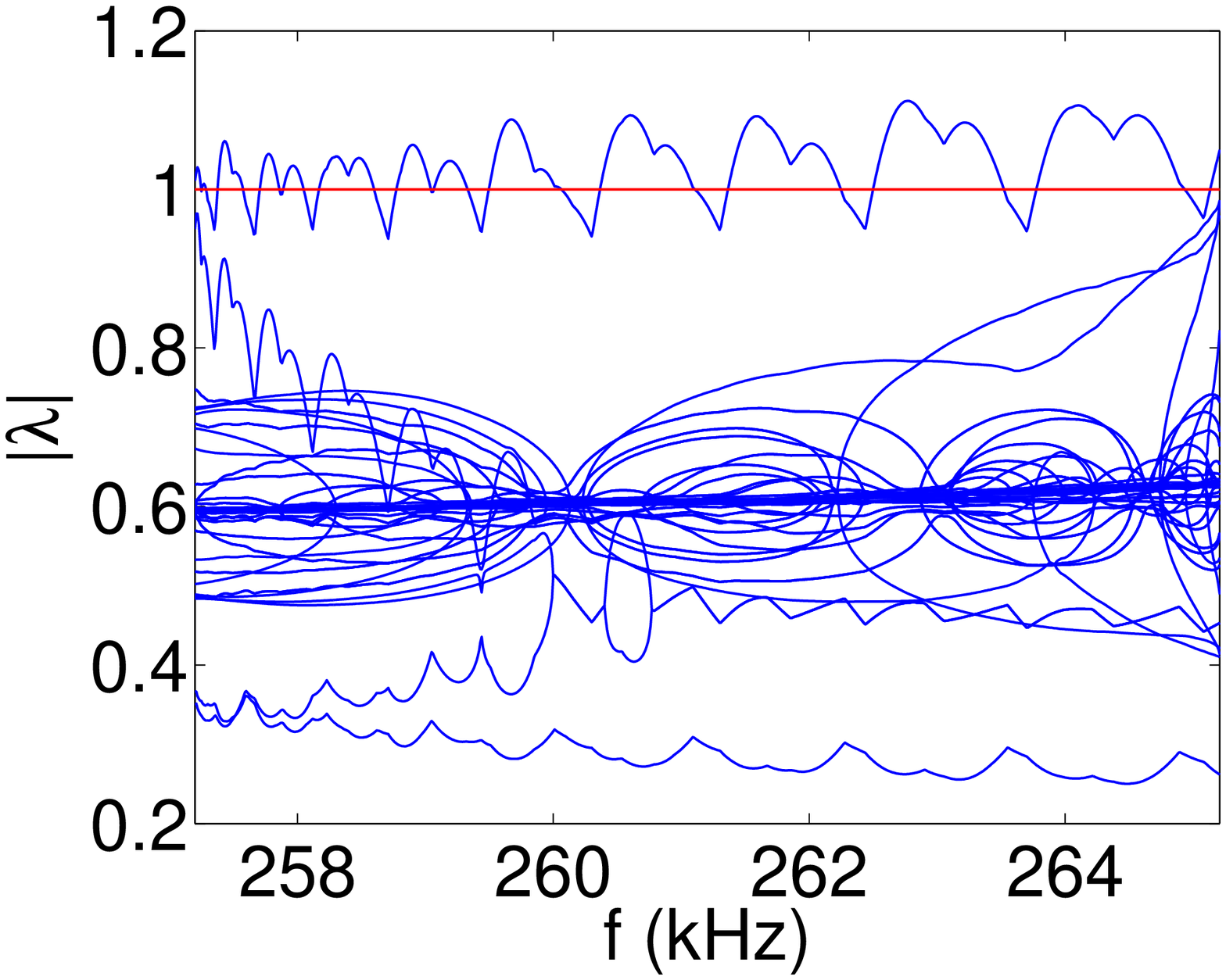} \\
\end{tabular}
\end{center}
\caption{(Color online).
{Same as Fig. \ref{exchange1} but for $V_d=2.5$~V.}}
\label{exchange2}
\end{figure}

\begin{figure}
\begin{center}
\begin{tabular}{cc}
\multicolumn{2}{c}{on-site breather} \\
\includegraphics[scale=0.23]{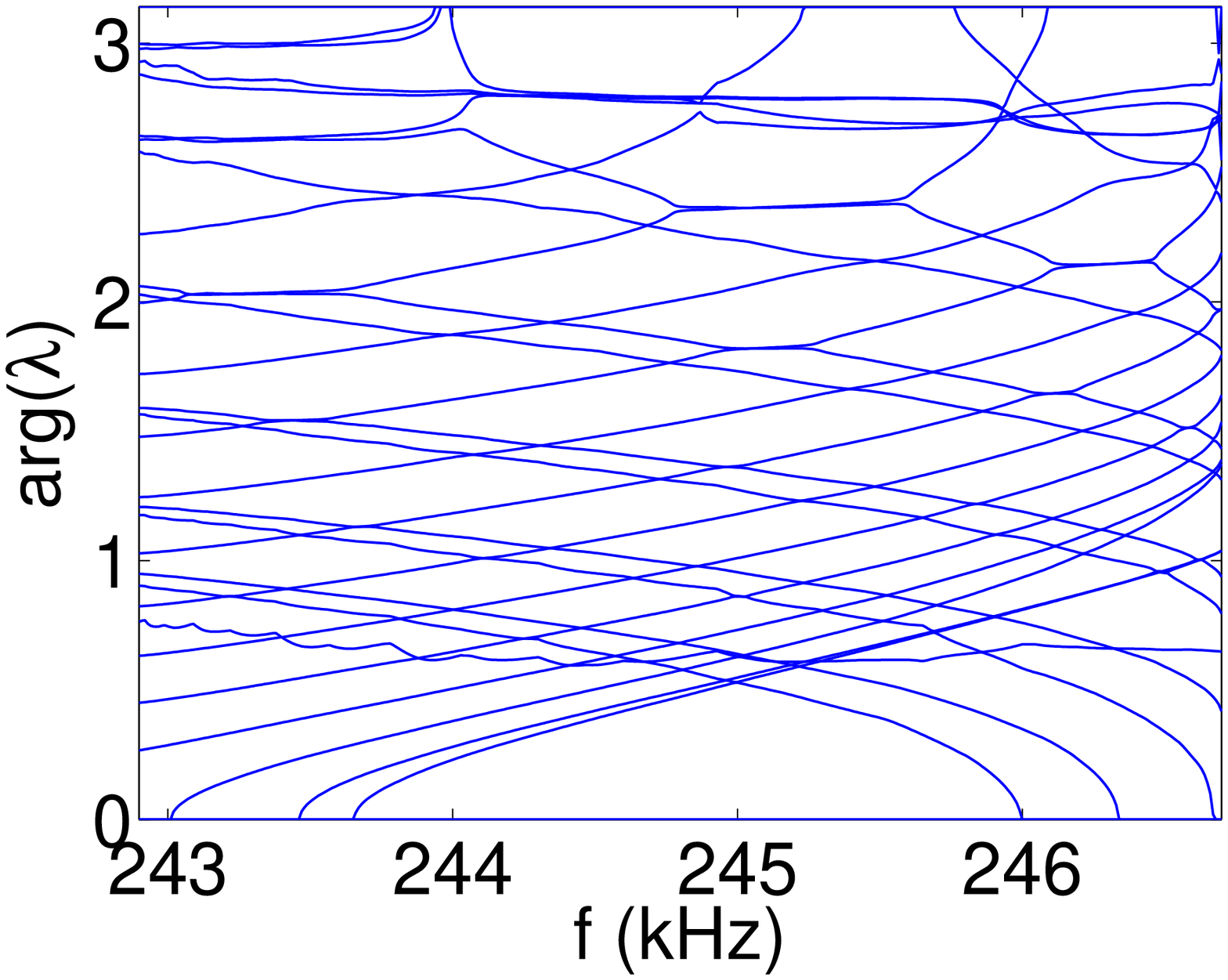} &
\includegraphics[scale=0.23]{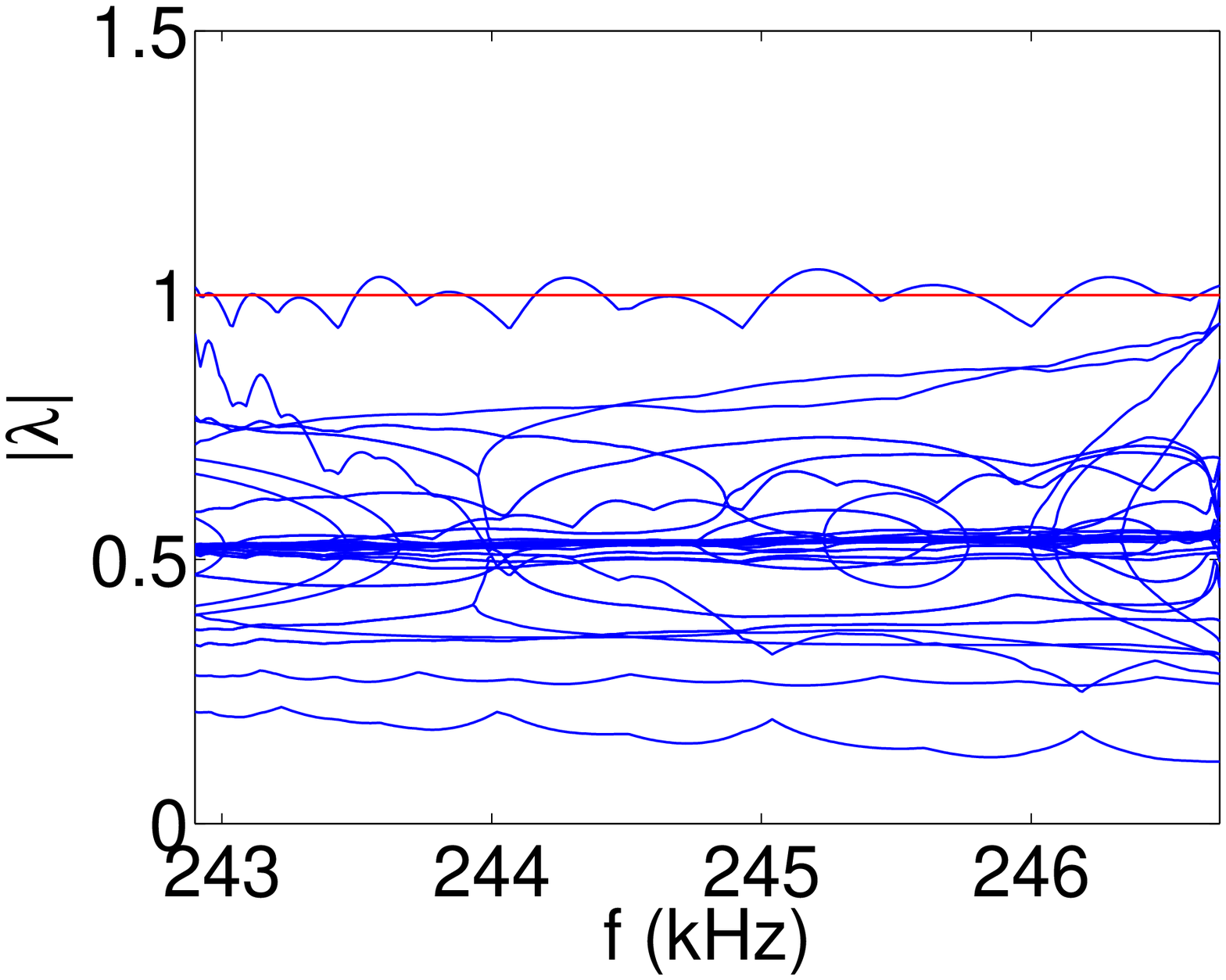} \\
\multicolumn{2}{c}{inter-site breather} \\
\includegraphics[scale=0.23]{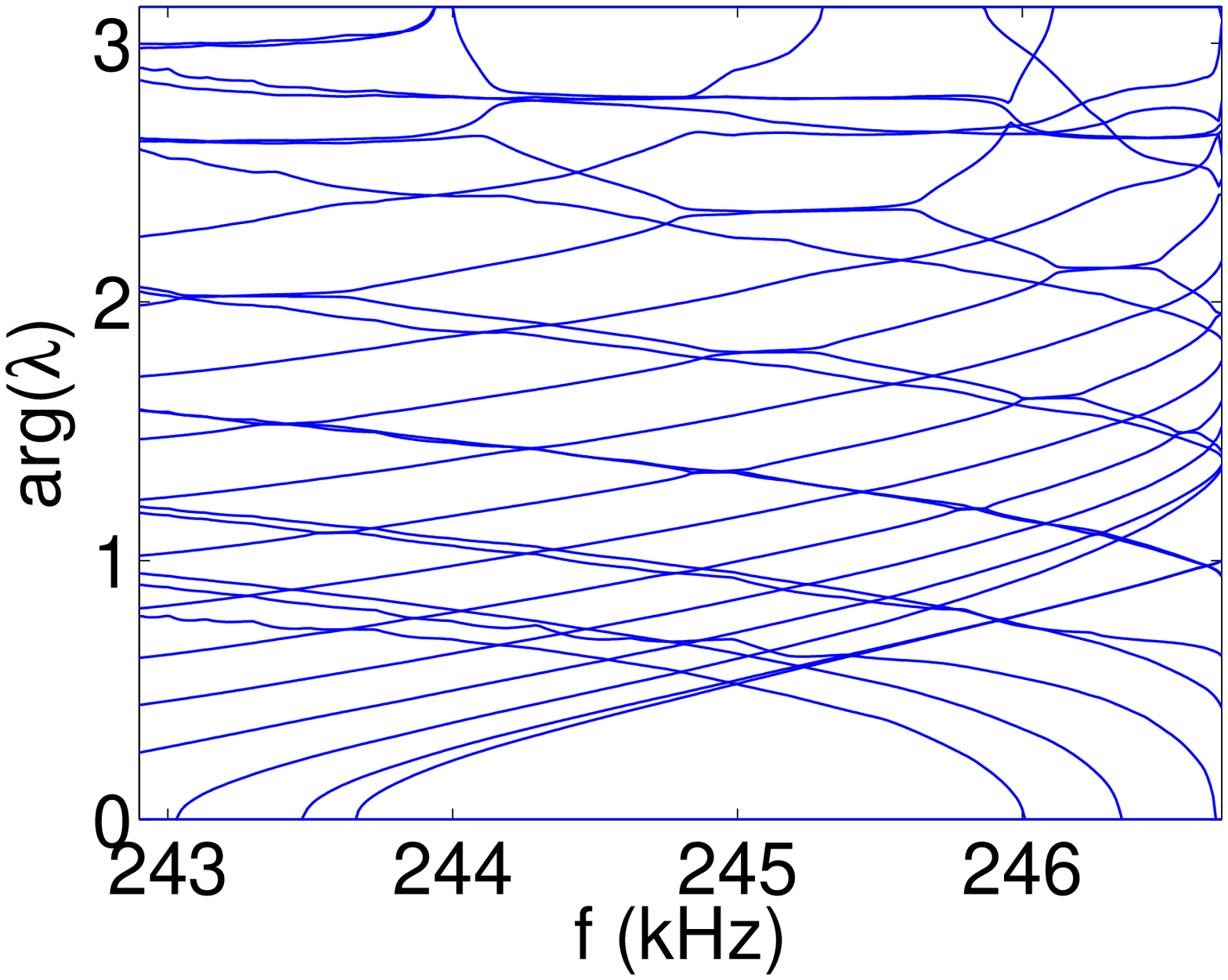} &
\includegraphics[scale=0.23]{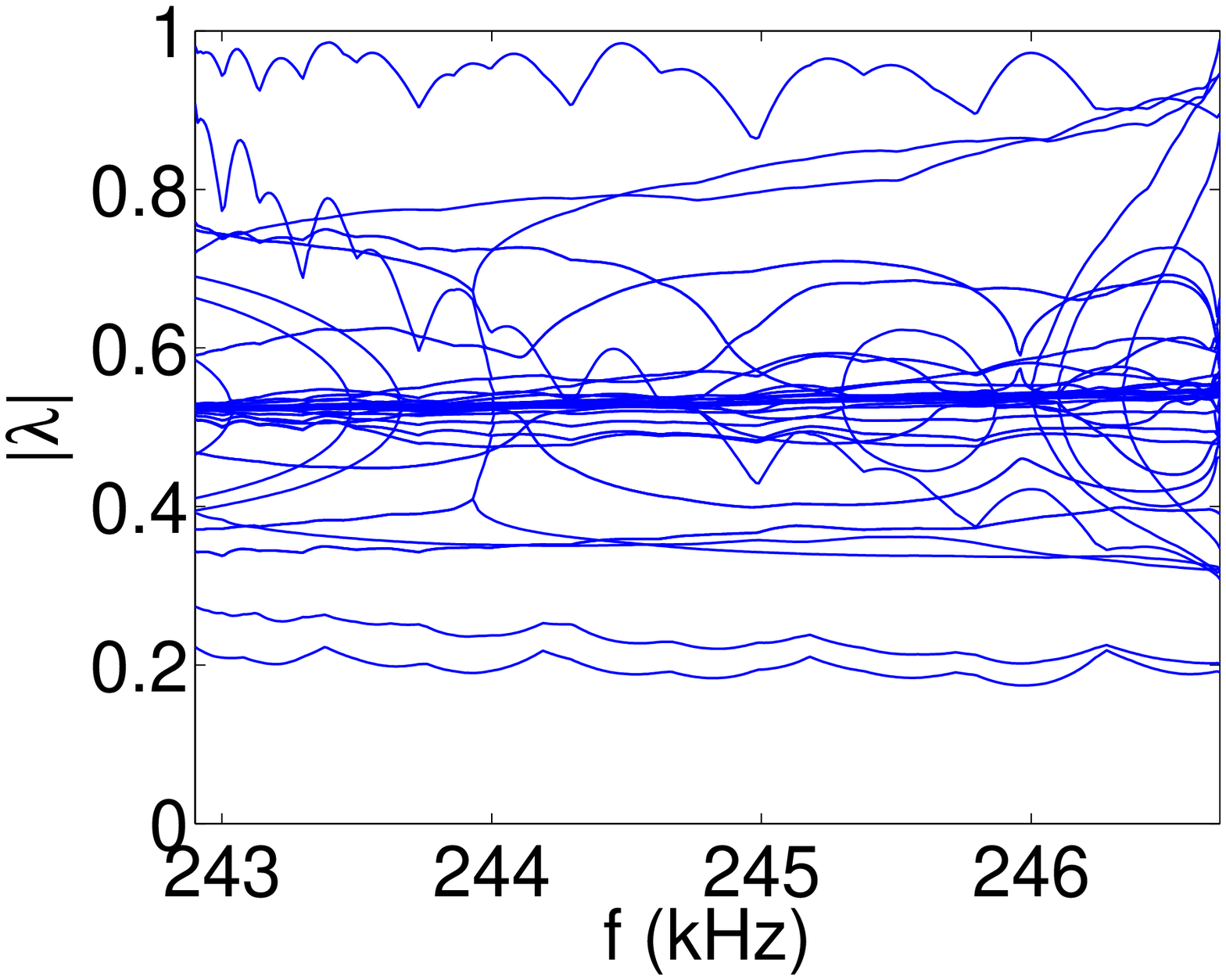} \\
\end{tabular}
\end{center}
\caption{(Color online).
{Same as Fig. \ref{exchange1} but for $V_d=4$~V.}}
\label{exchange3}
\end{figure}

A detailed analysis shows that for 
$3\lesssim V_d\leq 5$~V,
inter-site breathers are always stable, whereas on-site ones may be
unstable (i.e., there is no stability exchange as it occurs in
Klein-Gordon lattices). However, for $V_d\lesssim 2$~V, inter-site breathers are always unstable whereas on-site breathers are stable. Finally, for $2$~V$\lesssim V_d\lesssim 3$~V, inter-site breathers experience instability windows whereas their on-site counterparts are stable. 
Figures~\ref{exchange1}--\ref{exchange3} illustrate a typical set of 
relevant results for the cases of $V_d=1.5$~V, $V_d=2.5$~V and $V_d=4$~V
which summarize the different possible regimes as $V_d$ is varied.
The unity crossings indicated by the red line (wherever 
relevant) in the right panel
of each of the figures mark the stability changes of the pertinent solutions.

We have also performed an analysis of the stability of 1-peak breathers for larger lattices ($N=101$) and observe that the existence and stability range for on-site and inter-site breathers are not significantly altered.


To corroborate the numerical picture, we have also performed an analogous experimental analysis. To determine experimentally the
stability islands, we have chosen a particular voltage $V_d$ and, starting at low frequencies (homogeneous state), and increasing $f$ adiabatically, we can reach different regions. Going up and down adiabatically we are able to examine regions of coexistence between different $n$-peak breathers. In particular, regions corresponding to 1-peak, 2-peak and 3-peak breathers have been detected experimentally. The 3-peak region is bounded from above by the 4- and 5-peak areas, which we did not track. We have determined the boundaries by doing up-sweeps and down-sweeps, with significant hysteresis phenomena.

\begin{figure}
\begin{center}
\includegraphics[scale=0.4]{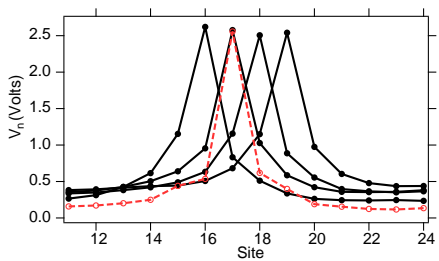}
\end{center}
\caption{The ILM can exist at any site of the lattice; here f = 253 kHz and $V_d$ = 4V. We demonstrate that the ILM can be made to jump to the neighboring site upon the temporary creation of an impurity there. The left-most profile (centered at site 16) turns  into the dotted (red) trace when an external inductor is placed in the direct vicinity of the $L_2$ inductor at site 17. Upon removal of this external inductor, the ILM persists at site 17 (solid black trace). This process can be repeated to move the ILM further to the right, as shown.}
\label{hop}
\end{figure}

It should be mentioned that for the interior of the 1-peak region, we demonstrated experimentally that the discrete breather/ILM can be centered at any node in the lattice and survive there. This verification implies that the experimental lattice ---despite its inherent component variability--- does display a sufficient degree of spatial homogeneity for the basic localization phenomenon to be considered (discrete) translationally invariant. In practice, we employed an impurity in the form of an external inductor physically touching a $L_2$ lattice inductor to make the ILM hop to that impurity site; see Fig.~\ref{hop}. Upon removing the impurity, the ILM would then persist at that site. We believe that this technique
could prove extremely valuable towards the guidance and manipulation
(essentially, at will) of the ILMs in this system.

\begin{figure}
\begin{center}
\begin{tabular}{cc}
\includegraphics[scale=0.4]{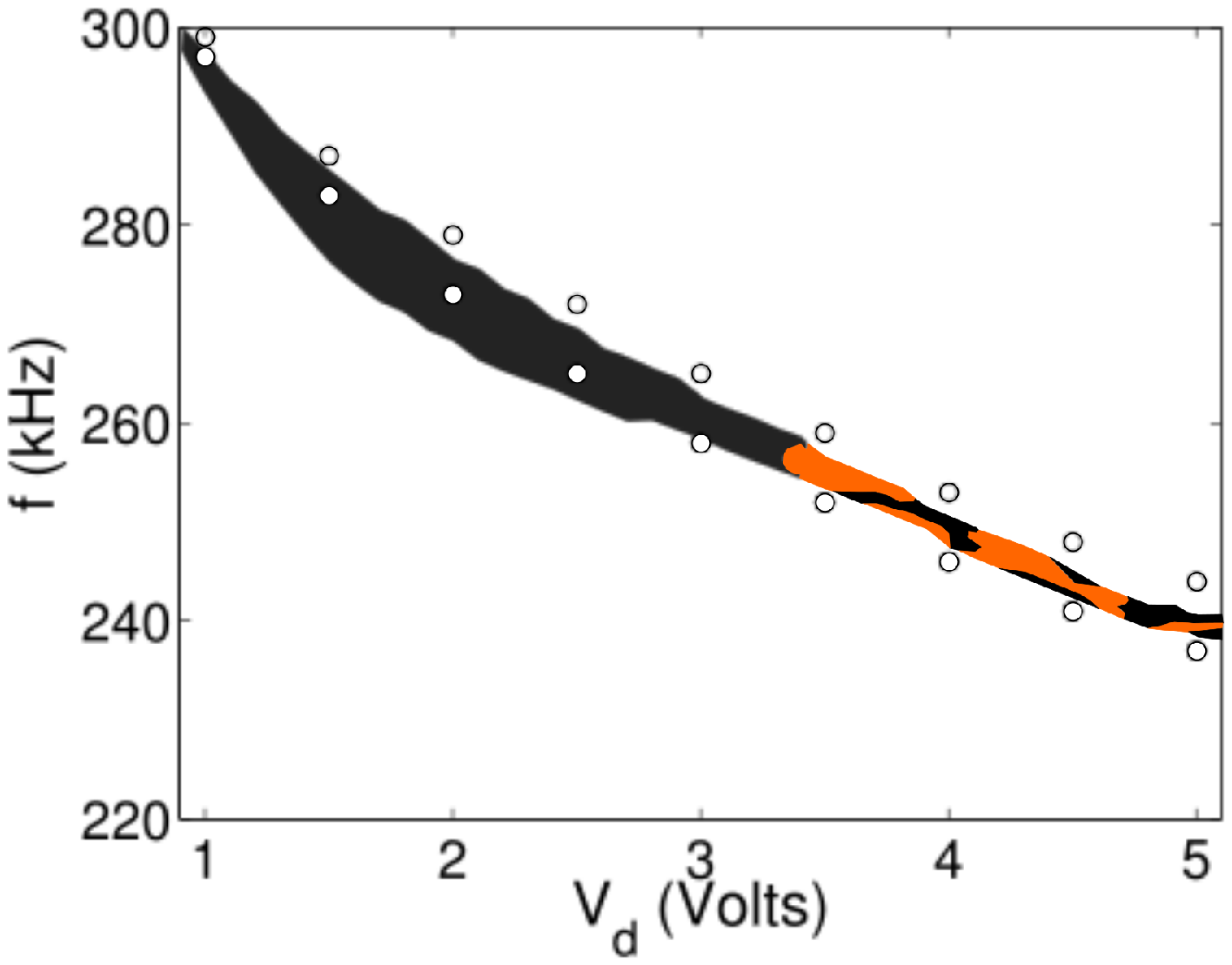} \\
\includegraphics[scale=0.4]{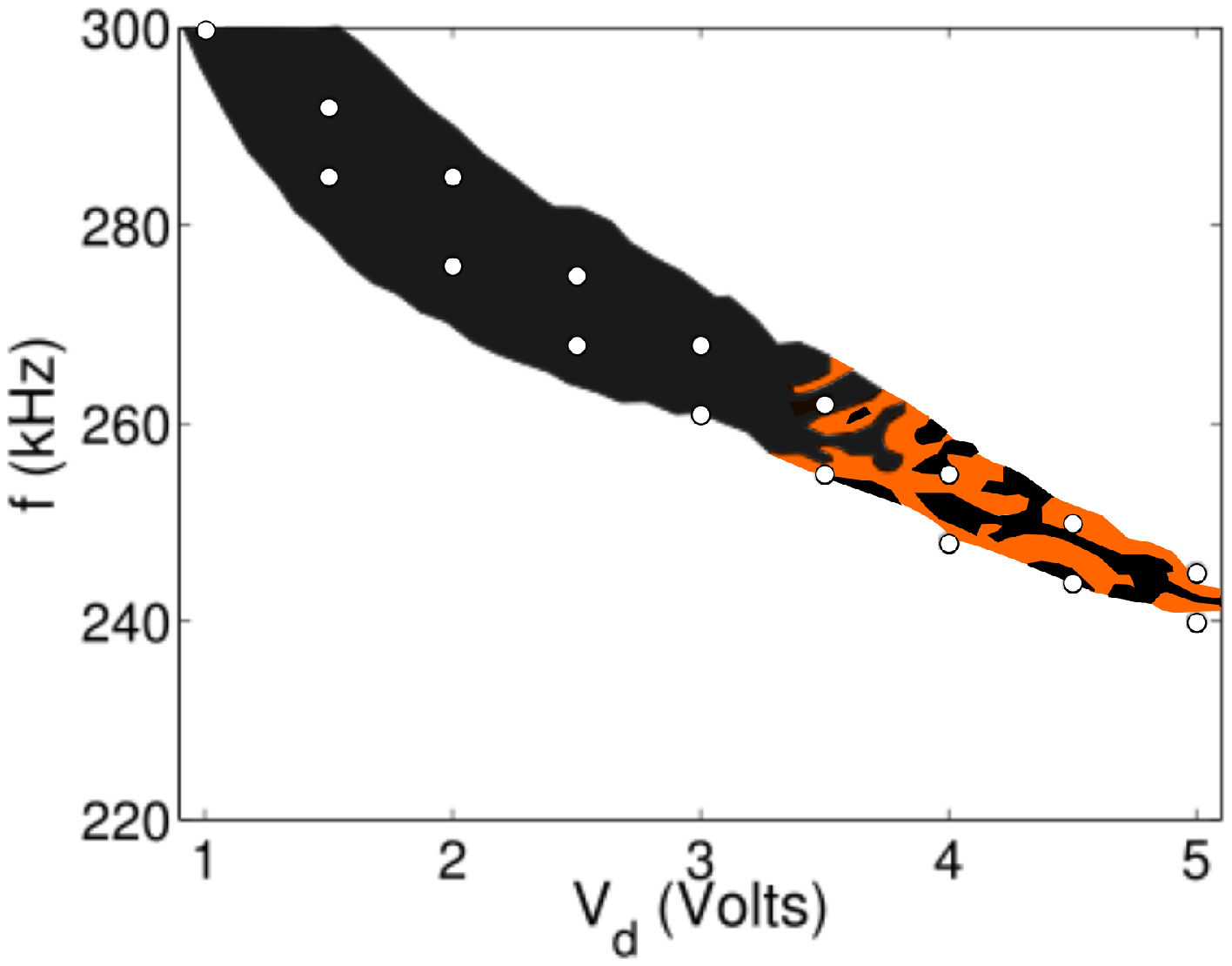}
\end{tabular}
\end{center}
\caption{(Color online).
Existence regions of 1-peaked (top panel) and a family of
3-peaked (bottom panel)  breathers
obtained numerically (black areas correspond to stable solutions and
orange (grey) one to unstable solutions) as compared to
experimental data identifying the range of observations of the corresponding
type of states (circles).
The theoretical data are displaced by a +7 kHz frequency offset.}
\label{comparison}
\end{figure}

A comparison between theoretical and experimental data of 1-peak and 3-peak
breathers is shown in Fig.~\ref{comparison}. The precise width of any
stability region is hard to match, because there exist both
regions of different peak
numbers and regions of different families with the same number of peaks
overlap, and thus their competition prevents an absolutely definitive
picture. Which one ``wins'' out appears to depend
sensitively on small lattice impurities present in our (commercial) experimental
elements. For instance, in the
experiment it looks as if the 2-peak region is squeezed in favor of
the 1-peak region. This might be the reason why the experimental 1-peak
region is slightly wider at higher driver amplitudes than
it appears in the corresponding theoretical predictions.
Nonetheless, the comparison between experimental and theoretical existence
regions depicted in Fig.~\ref{comparison} shows generally good
qualitative (and even quantitative) agreement in the context
of the proposed model.

\begin{figure}
\begin{center}
\begin{tabular}{c c}
~~$V_d=2$~V & $V_d=4$~V~~ \\[-0.5ex]
\includegraphics[scale=0.28]{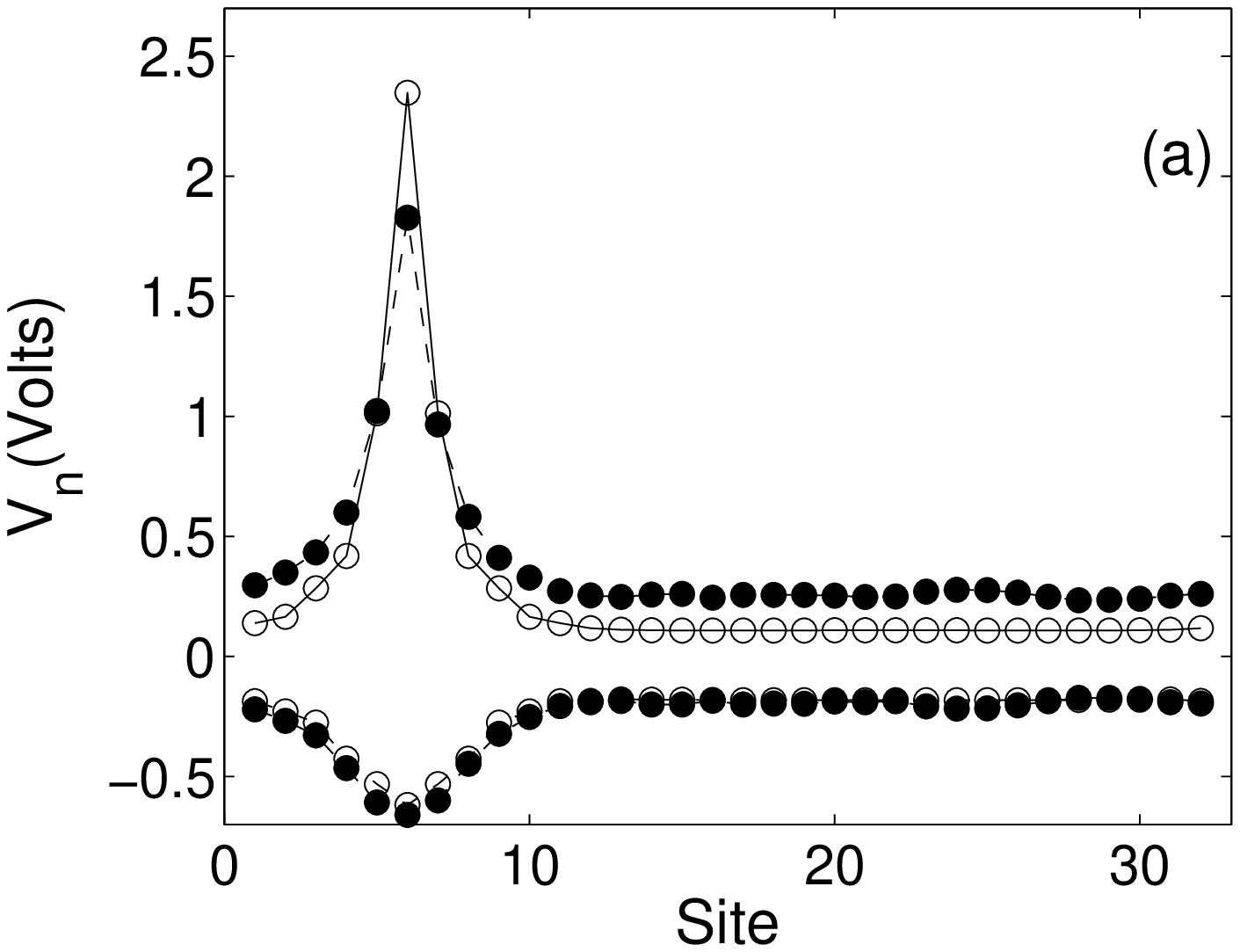} &
\hskip-0.45cm
\includegraphics[scale=0.28]{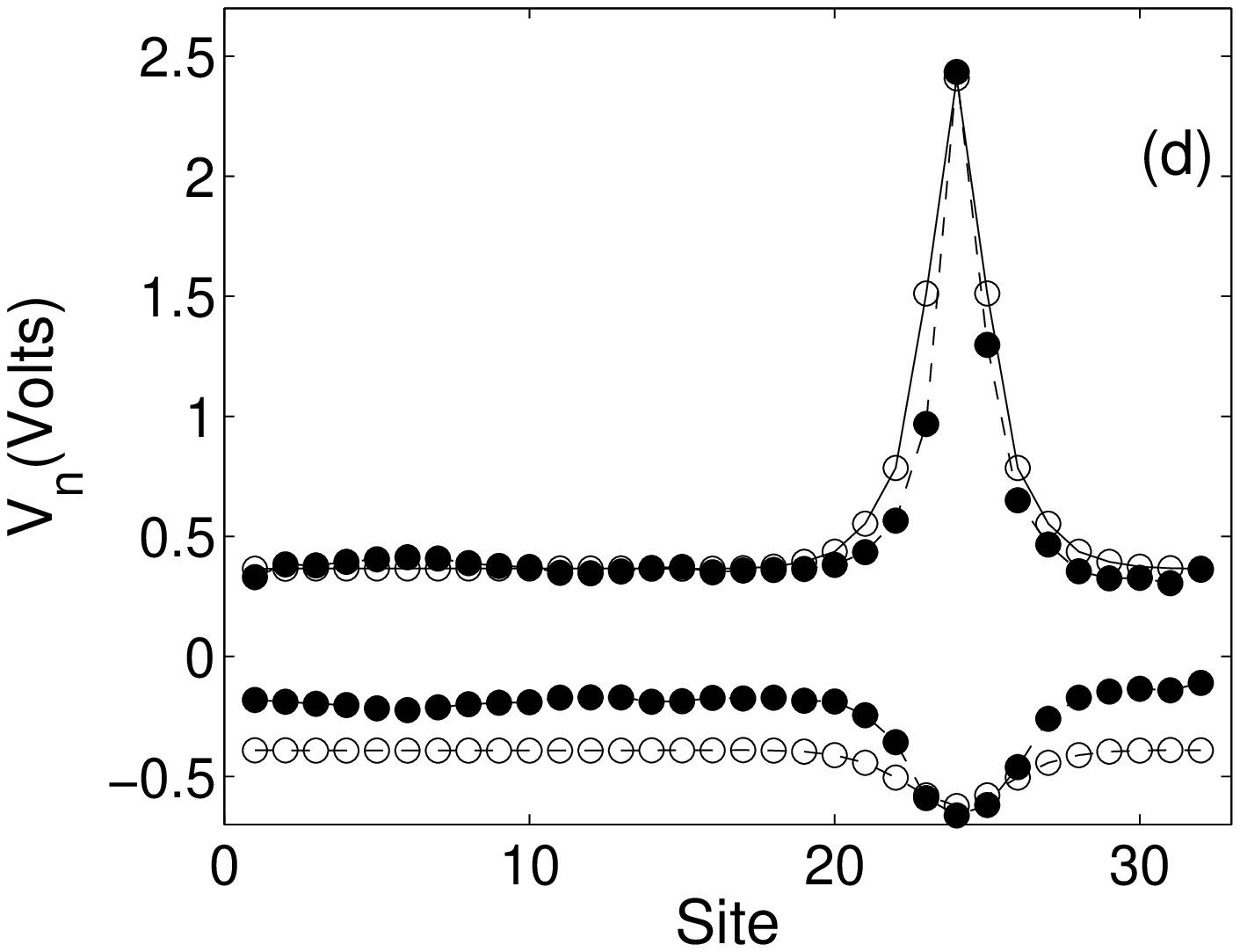} \\
\includegraphics[scale=0.28]{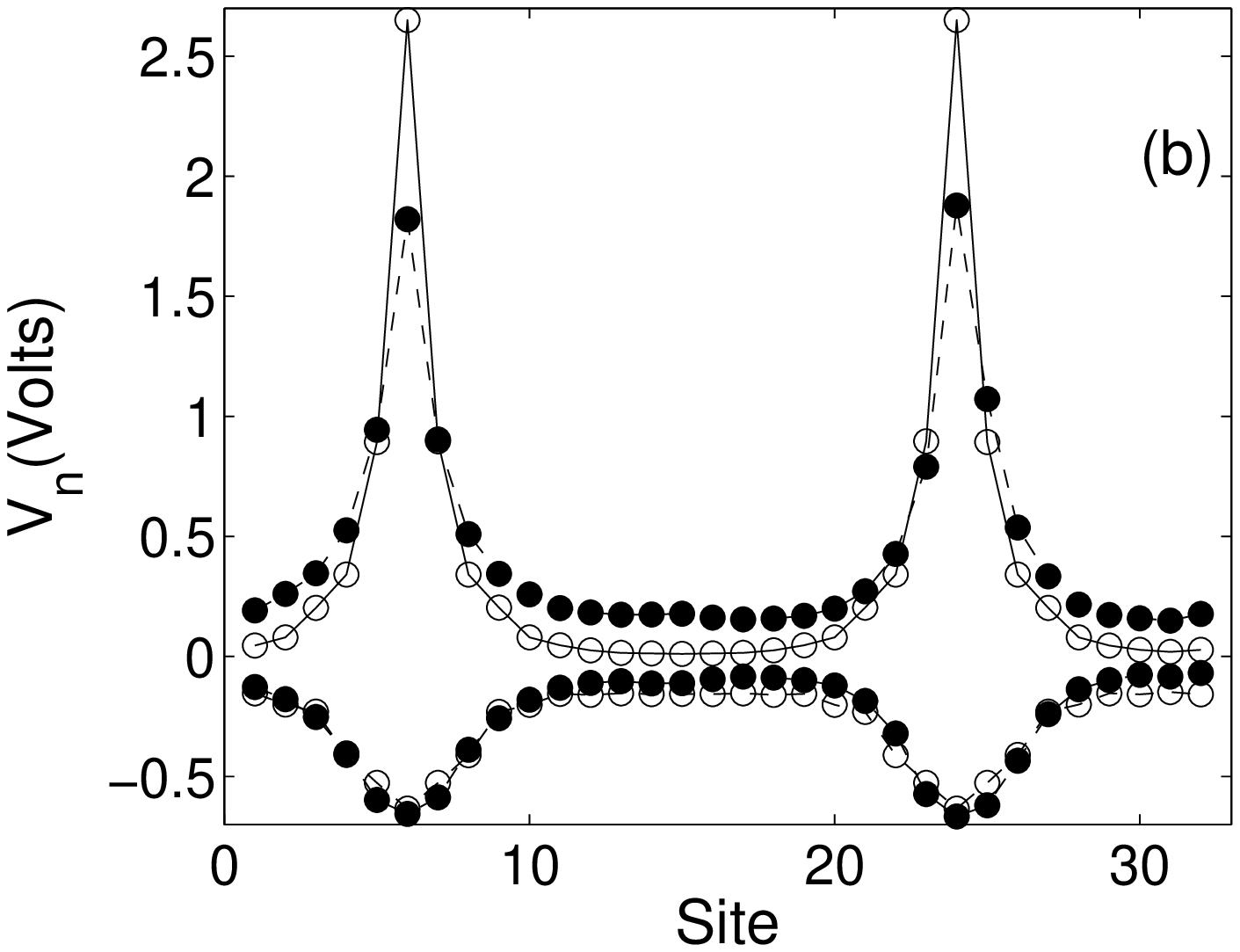} &
\hskip-0.45cm
\includegraphics[scale=0.28]{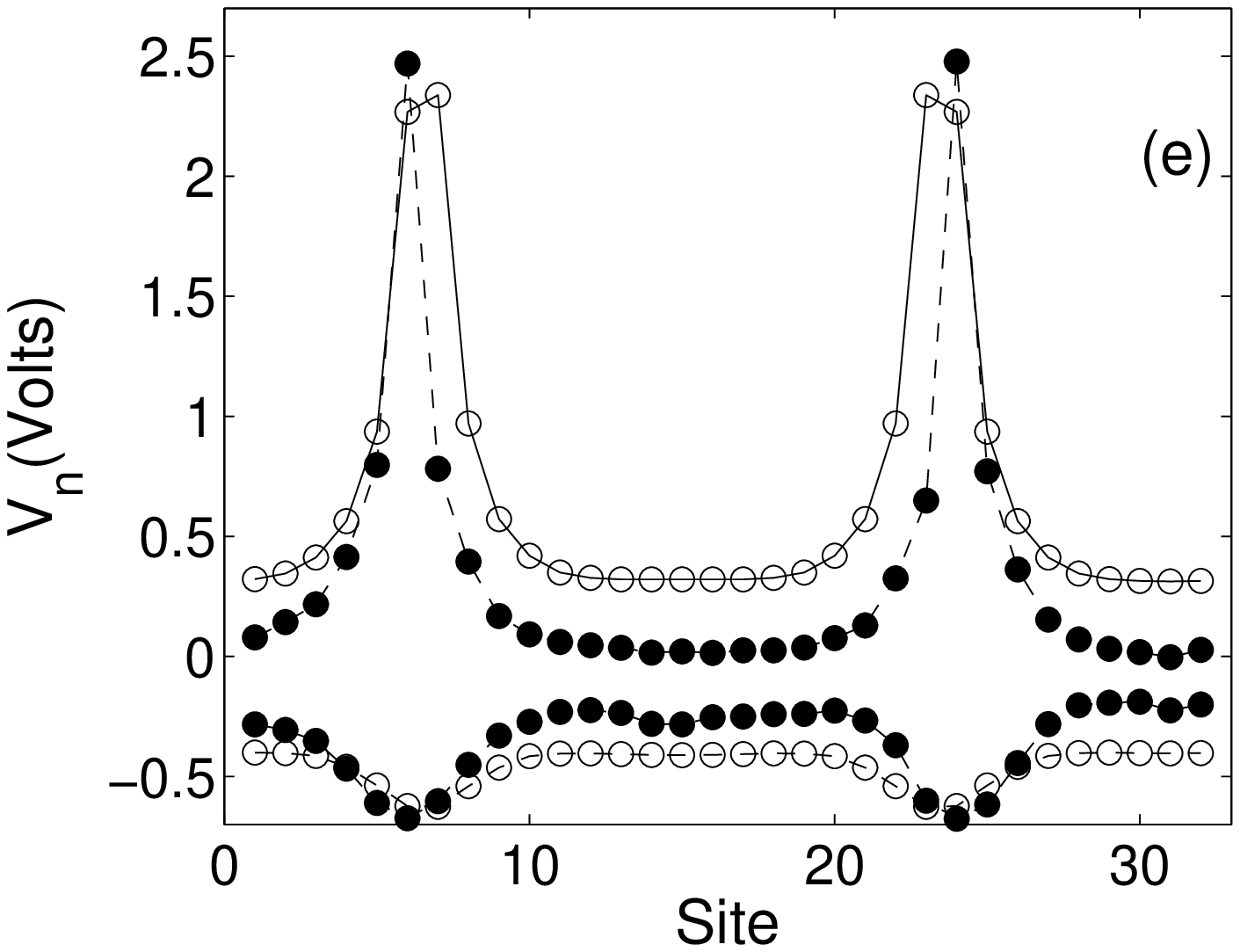} \\
\includegraphics[scale=0.28]{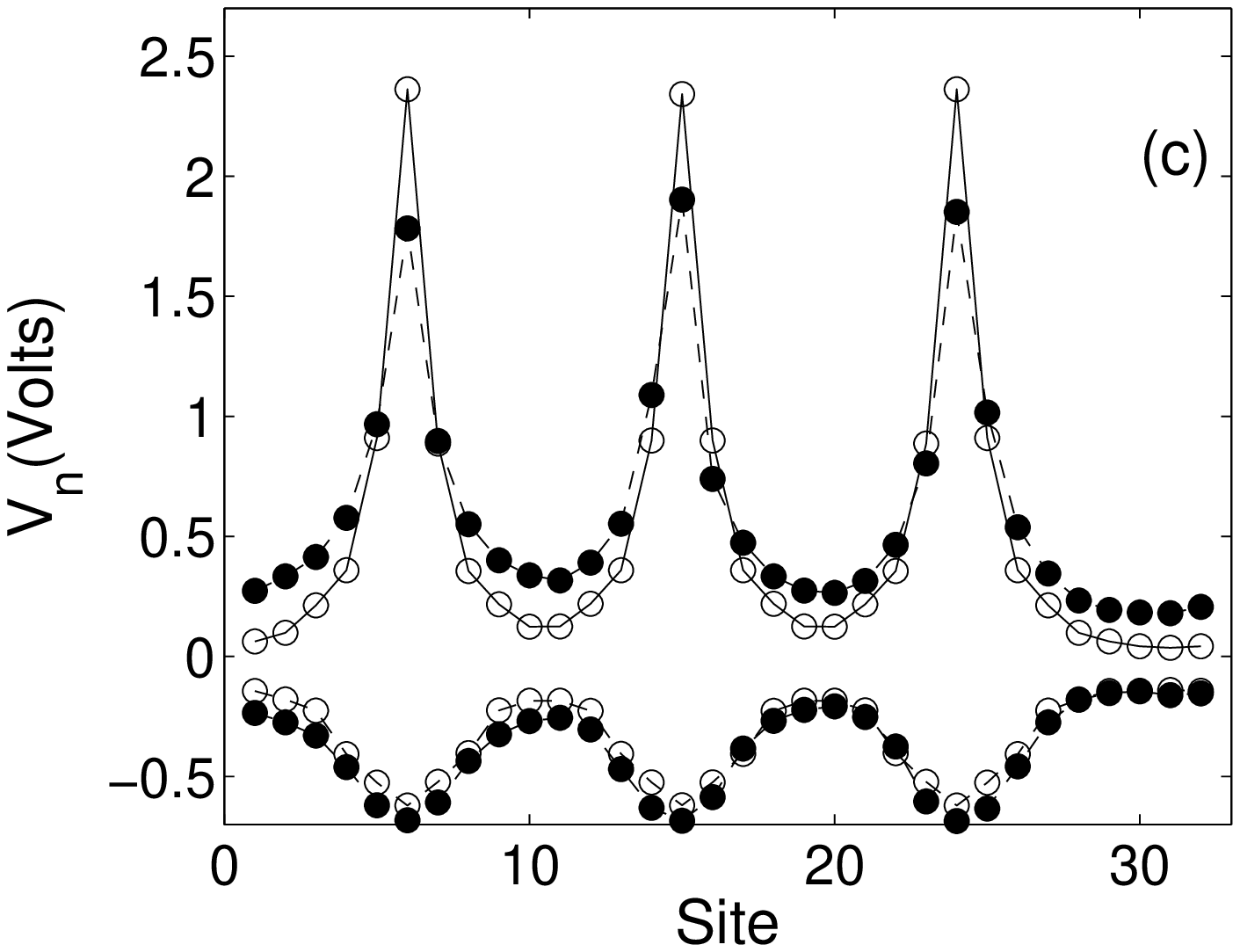} &
\hskip-0.45cm
\includegraphics[scale=0.28]{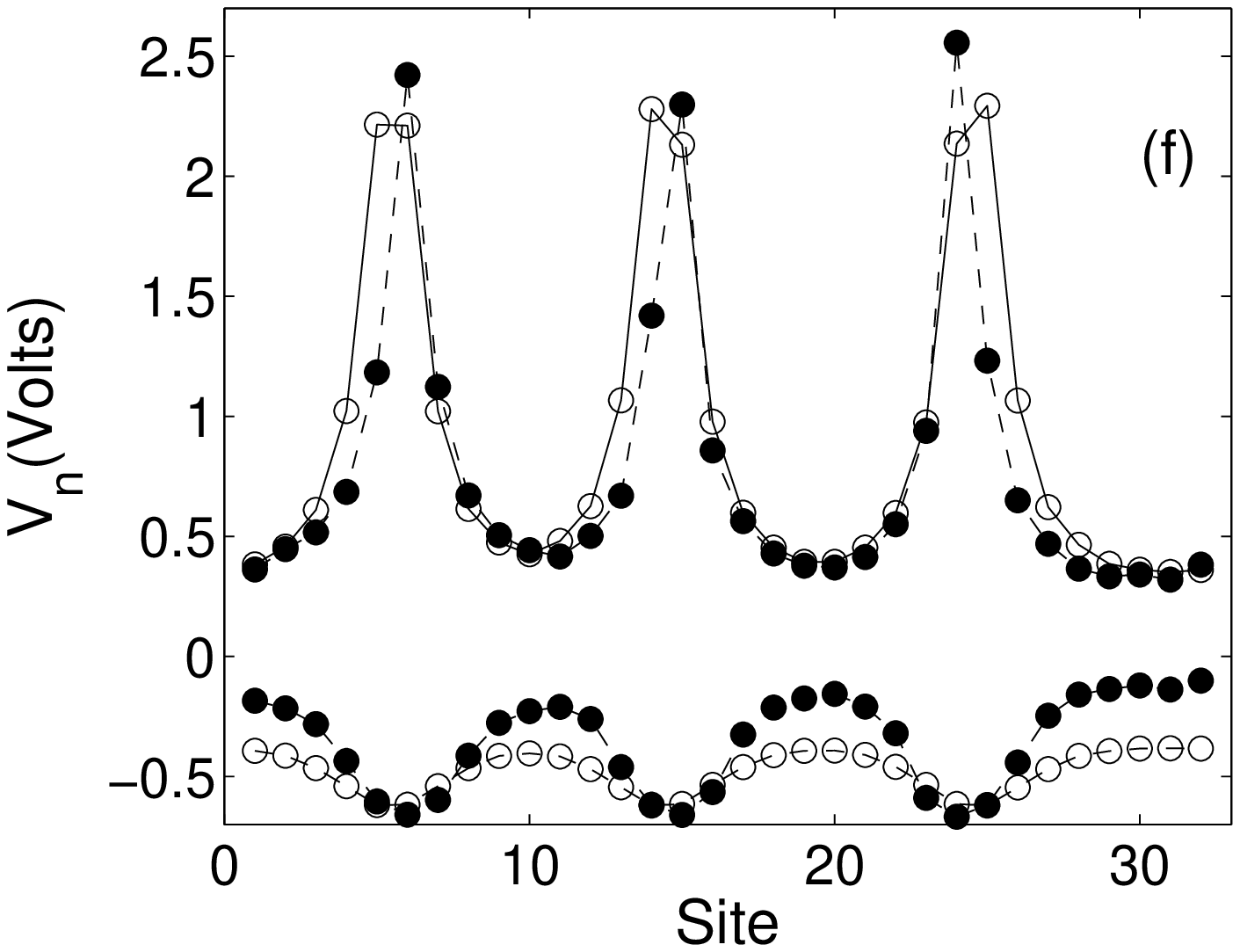}
\end{tabular}
\end{center}
\caption{Experimental ($\bullet$) and numerical ($\circ$)
1-, 2- and 3-peak breather profiles for different driver voltage
$V_d$ and frequencies values.
The left (right) column of panels corresponds to
a driver strength $V_d=2$~V ($V_d=4$~V).
Each panel correspond to the following
experimental ($f_{\rm exp}$) and numerical ($f_{\rm num}$) driving
frequencies in kHz:
(a) $(f_{\rm exp},f_{\rm num})$=(275,268),
(b) $(f_{\rm exp},f_{\rm num})$=(276,269),
(c) $(f_{\rm exp},f_{\rm num})$=(280,273),
(d) $(f_{\rm exp},f_{\rm num})$=(248,244),
(e) $(f_{\rm exp},f_{\rm num})$=(254,247) and
(f) $(f_{\rm exp},f_{\rm num})$=(249,245).
%
}
\label{profiles}
\end{figure}

More detailed experimental results are shown in Fig.~\ref{profiles}
where we depict peak profiles at two different driver voltages.
The profiles were taken at the times of largest peak voltage amplitude
and lowest peak voltage amplitude. For $V_d=2$V (left column of panels) we see that, as the frequency is raised from below, we cross from the 1-peak region through the 2-peak region and into the 3-peak region. For the $V_d=4$~V case (right column of panels) the same sequence can be observed when scanning in one frequency direction. In order to illustrate both the hysteresis and the overlap between $n$-peak regions, we depict in the figure (panels (e) and (f)) a situation where
the 2-peak solution occurs at a higher frequency than the 3-peaked one. The reason is that in Fig.~\ref{profiles}(f), the 3-peak solution was obtained at higher frequencies and then adiabatically extended to lower ones, whereas in Fig.~\ref{profiles}(e) the 2-peak solution was obtained starting from the 1-peak region.

We show the eventual location of peaks
in the breather pattern (i.e. after the driver has been on for a long time).
However, it is important to mention that the exact location
where the peaks eventually settle is sensitive to slight impurities in the
lattice. We have noticed that when we turn on the voltage source, at first
we can observe a more sinusoidal pattern (corresponding to the most
modulationally unstable
$k$-value), but as the pattern reaches higher energy and becomes more
nonlinear, the peaks may shift and adjust themselves in the lattice.
As it can be observed, peaks are not perfectly equispaced in the lattice. This is obviously due to the inhomogeneities and noise present
in the experiment.

For the numerical results depicted in Fig.~\ref{profiles}
we used a set of initials conditions based on the experimental data and
determined the stationary state by letting the numerical profiles
to settle to a steady state.
For the cases corresponding to $V_d=2$~V and $V_d=4$~V, adding a small
frequency offset $\Delta f\approx 4-7$~kHz, we observe, in general, a
good agreement between numerics and
experimental data. The mismatch between experiments and theory, in particular
the intersite distance peaks,
can be attributed to the above mentioned factors.
Furthermore, to reproduce precisely the experimental peak voltage is
extremely difficult because it corresponds to the voltage at resonance
and, therefore, even very small parameter changes can create large
differences in the maximum amplitudes.
Nevertheless, the quantitative agreement appears fairly good, especially for the $V_d=4$~V case.

\section{Conclusions}
\label{Sec:conclu}

In this paper we have formulated a prototypical model that is able
to describe the formation of nonlinear intrinsic localized modes
(or discrete breathers) in an experimental electric line.
This has been derived based on a combination of the fairly accurate
characterization
of a single element within the lattice (including its nonlinear
resonance curves and hysteretic behavior) and fundamental
circuit theory in order to properly couple the elements.
Comparison between theory and experiments shows very good
qualitative and even good quantitative agreement between the two.
%
We characterized the regions of existence and stability of $n$-peaked
breathers for $n=1,2,3$ and illustrated how transitions of the coherent
waveforms of one kind to those of another kind take place, rationalizing
them on the basis of stability properties and
their corresponding Floquet spectra.
We also showed that the precise number of peaks and their location in the
lattice is fairly sensitive to initial conditions, a feature also
generally observed in the experiments where the potential for states with different numbers of peaks similarly manifests.

Naturally, many directions of potential future research stem
from the fundamental modeling and computation basis explored
in the present manuscript. On the one hand, it would be very
interesting to attempt to understand the stability properties
of the different breather states from a more mathematical
perspective, although this may admittedly prove a fairly
difficult task. On the other hand, from the modeling and computation
perspective in conjunction with experimental progress, the present
work paves the way for potentially augmenting these systems
into higher dimensional setups and attempting to realize discrete
soliton as well as more complex discrete vortex states
therein~\cite{Flach2008,Kev09}. Such studies will be deferred
to future publications.

\medskip
\medskip
\medskip

\section{Acknowledgments}
F.P. and J.C. acknowledges sponsorship by the Spanish MICINN under
grant No.~FIS2008-04848.
R.C.G.\ gratefully acknowledges the hospitality
of the Grupo de F\'{\i}sica No Lineal (GFNL, University of Sevilla, Spain)
and support from NSF-DMS-0806762, Plan Propio de la Universidad de Sevilla,
Grant No. IAC09-I-4669 of Junta de Andalucia and Ministerio de Ciencia e
Innovaci\'on, Spain.
P.G.K.\ acknowledges the support from NSF-DMS-0806762,
NSF-CMMI-1000337 and from the
Alexander von Humboldt, as well as the Alexander S. Onassis Public
Benefit Foundation.

\end{document}